\documentclass[12pt,preprint]{aastex} 
\usepackage{enumerate}

\def\newhorizons{\textit{New Horizons}}
\def\Da{q}
\def\Damin{q_{_{\rm min}}}
\def\dm{\delta m}
\def\rclose{r_{_{\rm\!close}}}
\def\rroche{a_{_{\rm\!R}}}
\def\rxing{q_{_{\rm xing}}}
\def\rhill{r_{_{\rm\!H}}}
\def\vhill{v_{_{\rm H}}}
\def\vesc{v_{\rm esc}}
\def\vstab{v_{\rm Q}}
\def\vkep{v_{_{\rm K}}}

\def\tkep{T_{_{\rm\!K}}}
\def\tsynodic{T_{\rm syn}}
\def\tspread{T_{\rm spread}}

\def\tdamp{T_{\rm damp}}
\def\tmig{T_{\rm migrate}}

\def\tbin{T_{_{\rm\!PC}}}
\def\tcol{T_{\rm col}}

\def\fluxvisc{{J_{\rm visc}}}
\def\fluxgrav{{J_{\rm grav}}}
\def\ffoc{{f_{\rm foc}}}

\def\Mp{M_p}
\def\Ms{M_s}
\def\Rp{R_p} 
\def\Rs{R_s}

\def\MP{M_{\rm P}}
\def\RP{R_{\rm P}}

\def\MC{M_{\rm C}}
\def\RC{R_{\rm C}}

\def\aPC{a_{\rm PC}}
\def\ePC{e_{\rm PC}}
\def\iPC{i_{\rm PC}}
\def\abin{{a_{\rm bin}}}
\def\acrit{{a_{\rm crit}}}
\def\ebin{e_{\rm bin}}
\def\eforce{e_{\rm force}}
\def\efree{e_{\rm free}}
\def\nbin{{\Omega_{\rm bin}}}

\def\rgc{{R_{\rm g}}}
\def\ngc{{\Omega_{\rm g}}}
\def\nsyn{\omega_{\rm syn}}
\def\kappae{\kappa_{\rm e}}
\def\kappai{\nu_{\rm i}}

\def\floss{\bar{f}_{\rm loss}}
\def\flossv{f_{\rm loss}}

\def\nurad{\nu_{\rm rad}}

\def\rp{r}
\def\rpmin{{\rp_{\rm min}}}
\def\rpmax{{\rp_{\rm max}}}
\def\mp{m}
\def\rhop{\rho}
\def\vp{v}
\def\vv{u} 

\def\hipass{{\cal H}}

\def\vpLXXVcms{\left[\frac{\vp}{\textrm{\small 75\,cm/s}}\right]}
\def\rpIkm{\Bigg[\frac{\rp}{\textrm{\small 1\,km}}\Bigg]}
\def\rplargeXkm{\Bigg[\frac{r_{\rm large}}{\textrm{\small 10\,km}}\Bigg]}

\def\rpsmallIm{\Bigg[\frac{r_{\rm small}}{\textrm{\small 1\,m}}\Bigg]}
\def\aXXRP{\left[\frac{a}{\textrm{\small 20\,$\RP$}}\right]}
\def\aringXXRP{\left[\frac{a_{\rm ring}}{\textrm{\small 20\,$\RP$}}\right]}
\def\aPCVRP{\left[\frac{\aPC}{\textrm{\small 5\,$\RP$}}\right]}

\def\DaIIRP{\left[\frac{\Delta a}{\textrm{\small 2\,$\RP$}}\right]}
\def\SigmaXLcgs{\left[\frac{\Sigma}{\textrm{\small 40\,g/cm$^2$}}\right]}

\def\flossOpVI{\left[\frac{\floss}{\textrm{\small 0.6}}\right]}
\def\vvesc{\left[\frac{\vp}{\vesc}\right]}

\def\dadm{{\delta a_m}}

\def\dadtweakgrav{{\dot{a}_{\rm grav}}}

\def\Mring{{M_{\rm ring}}}
\def\Xring{{X_{\rm ring}}}
\def\aring{{a_{\rm ring}}}

\def\mreduced{{\mu}}

\def\lae{\lower 2pt \hbox{$\, \buildrel {\scriptstyle <}\over {\scriptstyle\sim}\,$}}
\def\gae{\lower 2pt \hbox{$\, \buildrel {\scriptstyle >}\over {\scriptstyle\sim}\,$}}

\begin{document}

\title{Evolution of a ring around the Pluto-Charon binary}

\author{Benjamin C. Bromley}
\affil{Department of Physics \& Astronomy, University of Utah, 
\\ 115 S 1400 E, Rm 201, Salt Lake City, UT 84112}
\email{bromley@physics.utah.edu}

\author{Scott J. Kenyon}
\affil{Smithsonian Astrophysical Observatory,
\\ 60 Garden St., Cambridge, MA 02138}
\email{skenyon@cfa.harvard.edu}

\begin{abstract}


We consider the formation of satellites around the Pluto-Charon
binary.  An early collision between the two partners likely produced
the binary and a narrow ring of debris, out of which arose the moons
Styx, Nix, Kerberos and Hydra. How the satellites emerged from the 
compact ring is uncertain.  Here we show that a particle ring
spreads from physical collisions and collective gravitational
scattering, similar to migration.  Around a binary, these processes
take place in the reference frames of `most circular' orbits, akin to
circular ones in a Keplerian potential. Ring particles damp to these 
orbits and avoid destructive collisions.  Damping and diffusion also
help particles survive dynamical instabilities driven by resonances 
with the binary. In some situations, particles become trapped near 
resonances that sweep outward with the tidal evolution of the
Pluto-Charon binary.  With simple models and numerical experiments, we
show how the Pluto-Charon impact ring may have expanded into a broad
disk, out of which grew the circumbinary moons.  In some scenarios,
the ring can spread well beyond the orbit of Hydra, the most distant
moon, to form a handful of smaller satellites. If these small moons
exist, \newhorizons\ will find them.

\end{abstract}


\section{Introduction}

With its rich system of satellites, the Pluto-Charon binary seems a
miracle of planetary dynamics.  Separated by about 17 Pluto radii 
\citep[1\,$\RP\approx 1200$~km;][]{young1994,young2007}, Pluto and 
Charon have masses of $\MP\approx 1.3\times 10^{25}$\,g and 
$\MC\approx 1.5\times 10^{24}$\,g \citep{christy1978,buie2006}.
Their moons are not far away: Styx, Nix, Kerberos and Hydra are all 
packed between 37~$\RP$ and 60~$\RP$, on coplanar, nearly circular 
orbits with periods that are close to 3:4:5:6 resonances with the 
binary \citep{weaver2006, showalter2011, showalter2012, showalter2013,
buie2013, brozovic2015, showalter2015}.  The total mass of
the satellites is a small fraction ($\lesssim$ 0.02\%) of the
binary's mass, roughly $3\times 10^{19}$~g to $3\times 10^{21}$~g
\citep{buie2006, brucker2009, showalter2011, showalter2012, youdin2012,
brozovic2015}.  Resonances with the binary and the gravitational
effects of the moons on each other keep the satellite system on the
verge of chaotic disintegration \citep{tholen2008, suli2009,
  winter2010, peale2011, youdin2012, showalter2015}.  How the 
satellites arrived at their fragile orbits around Pluto and Charon 
is a mystery.

This delicate orbital architecture makes the Pluto-Charon system a
challenge for theories of circumbinary planet formation
\citep[e.g.,][]{mori2004, quintana2006, pierens2007, raf2013, 
bk2015tatooine}. The
most promising scenario for the origin of the binary itself is that
two dwarf planets had a grazing collision, leaving them intact but
bound at a small separation of 4--5~$\RP$ \citep{mckinnon1989,
  stern1992, canup2005, canup2011}. Pluto and Charon then tidally
evolved, moving out to their present positions, locked synchronously
together with a 6-day orbital period 
\citep{farinella1979,dobro1997,peale1999,cheng2014b}.  The impact likely
produced a dynamically hot ring, confined to a region no more than a
few times the binary's initial separation, inside the present-day
orbit of Styx, the innermost satellite 
\citep{canup2005,canup2011}. 

In the impact scenario, the four moons come from the debris fragments
\citep{stern2006, canup2005}.  After settling at orbital distances of
5--30~Pluto radii \citep{canup2005}, the debris circularizes into a
narrow circumbinary ring.  Over time the ring spreads radially
outward, and small debris particles within it grow through
coagulation. The result is a satellite system that has expanded
outward by a factor of three or more in orbital distance
\citep{kb2014}.  A major uncertainty in this picture is how the ring
--- either with fully formed satellites or smaller debris --- can
spread to the moons' current positions. A compelling idea is that the
moons formed quickly and migrated outward, trapped in resonances that
expanded as the binary tidally evolved \citep{ward2006}. However, this
mechanism depends on details of the tidal evolution, and no single
model can place all of the moons in their present orbits
\citep{lith2008b, cheng2014a}.

Despite difficulties with the \citet{ward2006} resonant migration
scenario, it has an important implication. Even if satellites were
placed by hand in their present orbits just after the giant impact,
then at least some would be lost when unstable resonances swept
outward along with the expanding binary
\citep{lith2008b,cheng2014a,cheng2014b}. Collisional damping and
diffusion might mitigate the effects of the resonances, but only if
satellites evolve in partnership with a reservoir of smaller particles
to damp them \citep{walsh2015}. Therefore, the key to putting the
moons into their current orbits is likely coordination between growth
of the satellites, the depletion of the smaller debris, the spreading
of the ring, and the tidal evolution of the binary.  The primary
motivation for this work is to understand this interplay.

Here, with the Pluto-Charon system in mind, we investigate
circumbinary ring dynamics to see how a compact ring of growing
particles evolves. Through analytic estimates and numerical
calculations, we examine the role of particle viscosity
\citep{cook1964, gold1978} and gravity \citep{lin1979b, ward1997}. We
highlight the dependence of the spreading rate on the state of the
ring particles, particularly their random velocities within the ring,
and we consider the effects of {\it in situ} satellite growth
\citep{kb2014}. In developing simple parameterized models, we offer
pathways from an initially compact ring around a short-period binary
to the more extended system of moons we see today. Some models predict
the formation of more small moons at orbital distances beyond Hydra.

In this paper, we outline the main physical phenomena that are
important to the formation of a circumbinary satellite system like
Pluto-Charon's (\S2). Then we give a general introduction to ring
dynamics (\S3), including the effects of a central binary on the ring
(\S4).  Turning our analysis to Pluto-Charon, we describe models of
ring evolution (\S5), and conclude with some predictions for
\newhorizons\ (\S6), now rapidly approaching the Pluto-Charon system
\citep{stern2008}.

\section{Overview and Context}

\begin{figure}[htb]
\centerline{\includegraphics[width=7.0in]{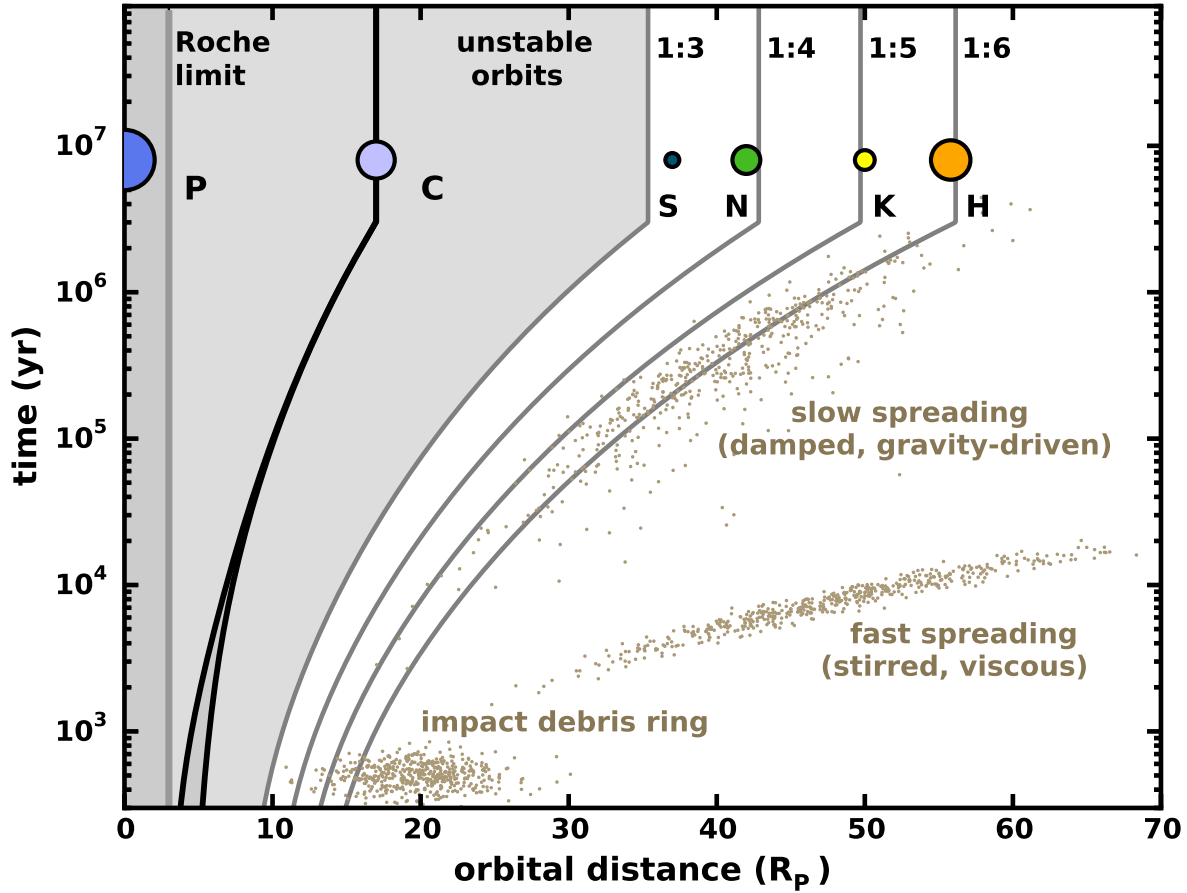}}
\caption{\label{fig:pcevolve} 
A schematic illustrating the layout and evolution of the
Pluto-Charon satellite system. The binary is on the left.
The satellites appear at their orbital distance from the
center of mass, with symbol sizes scaled according to
their approximate radii. The region inside of 3$\RP$ is the 
Roche zone. Within this region, particles are held together by 
their own material strength; collisions dominate the particle
dynamics. The light gray shaded region designates the inner cavity,
where most planar orbits are unstable. The gray curves extending to 
large orbital distances locate commensurabilities (as labeled) as 
the binary expands. The brown dots at the lower left indicate the
ring of debris after the binary formed, while the other clusters
of brown dots refer to fast and slow modes of ring spreading.
}
\end{figure}

The formation of the Pluto-Charon satellite system is a challenging
problem involving a range of physical phenomena.  The grazing
collision scenario for the formation of the binary yields only a
narrow, compact ring that somehow spreads, either from interactions
between ring particles or with the binary. The time scale for these
processes is uncertain. In some circumstances, the ring spreads rapidly
through gravitational stirring and physical collisions. In other
situations, the ring spreads slowly as a result of collisional damping
with only long-range, weak gravitational interactions to drive the
expansion.  In both cases, resonance excitations from the binary sweep
through the system and potentially destabilize satellite orbits.
One of our primary goals is to understand how the ring expands as the 
orbit of the binary evolves.
Figure~\ref{fig:pcevolve} illustrates the main issues, along with the
overall layout of the Pluto-Charon system. The following list provides
more detail about characteristics of the binary, the ring around it,
and the physical processes that guide the emergence of the satellite
system.

\begin{itemize}

\item
\textbf{The circumbinary environment: the inner cavity.} The
Pluto-Charon binary provides a time-varying potential that complicates
the nature of orbiting satellites \citep{lee2006}. Circumbinary
orbits near Pluto and Charon are largely unstable, leaving an inner
cavity that is about twice the binary separation, just inside the
orbit of Styx \citep[e.g., the light gray shaded region in 
  Figure~\ref{fig:pcevolve}, see also][]{holman1999, musielak2005, 
  pichardo2005, doolin2011}.  Carved out by instabilities at 
overlapping resonances \citep{wisdom1980} and by resonant-driven
torque exchange \citep[e.g.,][]{gold1980, meyer1987, ward1997},
this cavity sets the inner boundary of the circumbinary ring.  
Due to uncertainties in this complex environment, we consider two 
approaches to the boundary: (i) loss of ring material by ejection 
and (ii) a balance between viscous inflow and the outward torque 
from Charon.

\item
\textbf{Resonances.}  Aside from defining the edge of the inner cavity,
unstable resonances lie within the inner cavity and (for eccentric
binaries) at isolated larger orbital distances \citep[e.g.,][]{ward2006, 
cheng2014b}.  Current analyses favor a circular Pluto-Charon orbit;
resonances beyond the cavity's edge at the 3:1 commensurability are 
stable (e.g., Figure~\ref{fig:mostcircpcx}, below). Initially, the
binary probably had significant eccentricity \citep{canup2005}; 
resonances at higher order commensurabilities, include those at
4:1, 5:1, and 6:1, were then unstable.  Here we explore the effect of 
these resonances in the context of ring dynamics (\S\ref{subsec:rez})
and show that sufficiently strong collisional damping stabilizes 
particles within these resonances.

\item
\textbf{Stable circumbinary orbits.} Despite the resonances, the
binary has little impact on much of the orbital domain outside the 
inner cavity. As demonstrated by analytic theory \citep{lee2006, 
leung2013} and numerical experiment \citep{bk2015tatooine}, 
{\it non-resonant orbits experience no secular excitation or long-term 
torque exchange with the binary} (\S\ref{sec:circumbin}).  Particles 
tend to follow trajectories that ebb and flow with the orbital motion
 of the binary. When swarms of particles experience orbital damping, 
they settle on a family of ``most circular'' paths that do not 
intersect, just like circular paths around a single central mass 
\citep{youdin2013}. For a reference frame tied to a most circular 
orbit, the local dynamics within a ring around a binary is very similar
to dynamics around a single central mass. After describing these 
features (\S\ref{sec:circumbin}), we take advantage of them when 
modeling ring evolution (\S\ref{sec:pc}).

\textbf{Ring stability considerations: total mass and particle size.}
We also consider the gravitational stability of a ring with velocity
dispersion $v$, orbital frequency $\Omega$, and surface density
$\Sigma$ orbiting Pluto-Charon. When ring particles achieve a balance
between gravitational stirring and collisional damping, the velocity
dispersion is roughly the escape velocity of the largest particles in
the ring.  Defining $G$ as the gravitational constant, the Toomre
stability criterion, $v\Omega>\pi G\Sigma$, establishes stable
configurations of ring parameters \citep[e.g.,][]{chiang2010}.
Adopting a ring radius of 20~$\RP$, width of 5~$\RP$ and mass,
$3\times 10^{20}$~g, comparable to the mass of the known satellites, 
the ring is stable if the velocity dispersion exceeds the escape 
velocity of a 10-meter icy particle, roughly 1~cm/s.  Unless mergers 
produce larger particles which gravitationally stir the smaller 
particles up to their escape velocity, a ring composed of smaller 
particles is unstable.  In most of the models discussed here, we 
assume a sea of particles with radii of 1~km; rings composed of these 
particles are stable up to masses of $4\times 10^{22}$~g, roughly 
two orders of magnitude more massive than the binary's satellites.

\textbf{Ring dynamics.} We envision the spreading of the ring from its
early compact configuration through particle interactions. When
particle velocities exceed their escape speed, either from stirring or
because they are in the Roche zone of the central mass, physical
collisions dominate the dynamics \citep[][]{ cook1964, gold1978}.
Then, an effective collision-driven viscosity drives the evolution of
the ring.  Outside the Roche zone (which is only about 3~$\RP$ for
Pluto) collisional damping reduces the relative particle speeds until
gravitational interactions are important \citep{horn1985, shu1985}. 
Here, we provide a theoretical prescription for how these longer range 
interactions cause the ring to evolve. As illustrated in
Figure~\ref{fig:pcevolve}, collisionally driven viscosity, with stirring 
by larger embedded particles, leads to fast migration. Purely
gravitational spreading is comparatively slow.

\item
\textbf{Evolution of the binary.} 
Tidal evolution of the binary lies at the heart of all these issues.
After the grazing impact, the binary has a tighter, more eccentric
orbit. Tides circularize and expand the system into the current
circular orbit \citep[e.g.,][]{cheng2014a}.  Our analysis considers
how tidal evolution drives a set of sweeping mean motion resonances
(e.g., Figure~\ref{fig:pcevolve}) which lead to the dynamical
ejection of satellites \citep{cheng2014b}. By including collisional
damping and ring viscosity, we begin to show how growing satellites 
or their precursors experience resonant trapping, allowing them to 
migrate outward with the binary's tidal expansion. Alternatively, 
if the ring spreads quickly, we consider how satellites might survive, 
or at least reform, after the resonances sweep past.
\end{itemize}

While the circumbinary nature of Pluto-Charon satellites make them
unique in the Solar System, key physical processes governing their
formation are also important to other Solar System objects.  For
example, the regular satellites of Jupiter grow within a viscously 
spreading circumprimary disk \citep[see][]{lun1982, canup2002, 
mosque2003a, mosque2003b, mosque2010, sas2010, ward2010, ogi2012}. 
Although gas dynamics is crucial to this evolution, our analysis 
addresses aspects of the spreading of large particles uncoupled 
from the gas.  With its gas free environment, the Pluto-Charon
satellite system is more akin to Saturn's rings
\citep[see][]{canup1995, porco2007, charnoz2010}. Despite clear
differences in the location of small particles relative to the 
Roche limit (e.g., mostly inside for Saturn and entirely outside
for Pluto-Charon), both systems are driven by pure particle dynamics.
To extend the theory of Saturn's rings to a ring of solids orbiting 
Pluto-Charon, we focus next on the physics of particle rings.

\section{Ring dynamics}\label{sec:ringdyn}

Planetary rings provide an excellent theoretical proving ground for
planet formation.  Saturn's rings famously became a subject for James
Clerk Maxwell, and inspired a powerful theory for evolution within the
Roche limit of the planet \citep{cook1964,gold1978}.  Inside this
radial distance from Saturn, tidal forces inhibit particle growth.
Thus, the theory concentrates on the radial spreading and velocity
evolution of small, indestructible particles with negligible
gravitational interactions. Beyond the Roche limit, rapid growth into
small moons is a likely outcome; radial spreading and gravitational
scattering within the rings remain secondary \citep{charnoz2010}.  For
a ring of particles orbiting Pluto-Charon outside the Roche limit,
gravitational interactions between large particles add to the
spreading from small particles. Furthermore, dynamical spreading and
particle growth occur simultaneously.  To understand the evolution of
this material, we require a more general treatment.

Here we establish a theory for describing the major influences on the
evolution of a planetary ring.  Following \citet{gold1978}, we discuss
the role of physical collisions under the assumption that the
particles rebound off one another as they collide, changing their
trajectories but not their masses or radii
(\S\ref{subsec:coll}). Inelastic collisions allow damping of random
speeds \citep[e.g.,][]{bridges1984} and convert orbital energy into
random motions, yielding an effective viscosity that spreads material
radially \citep[\S\ref{subsec:visc},
  below]{jeffreys1947,cook1964,gold1978}. We also consider the role of
mutual gravitational interactions. Gravitational scattering affects
the equilibrium velocities that enter in the viscosity
\citep{horn1985, shu1985, gold2004}, as well as the radial spreading
of particles from collective interactions similar to migration
\citep{lin1979b, gold1980, ward1997}. Finally, we give a prescription
for the evolution of a ring in which particles grow. By incorporating
all of these phenomena together, we construct simple models for ring
evolution and lay the foundation for including this physics into full
evolution codes \citep[e.g.,][]{weiden1997b,kl1998,kb2014}.

\subsection{Background} 

Disks or rings around a central mass are a natural outcome in many
astrophysical systems. Conservation of angular momentum, along with
dissipative processes (e.g., radiative cooling of gas or inelastic
collisions between particles) drive material toward these flattened
configurations \citep[e.g.,][]{lbp1974,brah1976,pri1981}. In this 
section, we examine a ring of particles around a central point pass 
in the limit where the self-gravity of the ring is not important.  We 
begin by establishing some basic physical scales and relationships 
between physical properties of the ring.

To model the ring, we first assume that it is made of identical
particles with fixed radius $\rp$, mass $\mp$, and density $\rhop$.
A particle at an orbital distance
$a$ has an orbital period
\begin{eqnarray}
\tkep & = & 2\pi\sqrt{a^3/GM},
\end{eqnarray}
and Keplerian velocity
\begin{eqnarray}
\vkep & = &  \sqrt{GM/a},
\end{eqnarray}
where $M$ is the central mass and $G$ is the gravitational constant.
We assume that the orbits of the ring particles are approximately 
circular and have instantaneous speeds close to $\vkep$.

For orbits near the central mass, tidal forces are strong.  The Roche
limit gives a critical value,
\begin{equation}
\rroche \approx 1.5 \left(\frac{M}{\rhop}\right)^{1/3},
\end{equation}
inside of which massive particles are unable to hold themselves
together by their bulk strength and self-gravity
\citep{holsapple2006,holsapple2008}.

When particles are well outside of the Roche limit, they 
interact gravitationally with other neighboring satellites
\citep{salmon2010,charnoz2011,rosen2012}.  The gravitational 
range is characterized by the Hill radius,
\begin{equation}
\label{eq:rhill}
\rhill = a\left(\frac{\mp}{3M}\right)^{1/3}.
\end{equation}
The particle's Hill velocity,
\begin{equation}
\label{eq:vhill}
\vhill  =  a\sqrt{G\mp/\rhill},
\end{equation}
is the speed of a low-mass neighbor orbiting the particle at the 
Hill radius. Neighbors passing by at much faster speeds are 
undisturbed; slower objects are stirred to at least this speed 
\citep[see][]{gold2004}.

The fastest speed to which a particle can stir a less massive neighbor is
approximately the escape speed at its surface \citep[see][]{gold2004}:
\begin{eqnarray}
\label{eq:vesc}
\vesc & = &  \sqrt{2G\mp/\rp}  = \sqrt{8\pi G\rhop\rp^2 / 3}.
\end{eqnarray}
During an encounter, if the relative speed of an object and the 
satellite is small compared to $\vesc$, gravity influences the outcome.  
When the relative speed exceeds $\vesc$, gravitational interactions 
are much less important than physical collisions. 

In a swarm of identical particles, the escape velocity provides an
estimate of the characteristic speed of particles, $\vp$, as measured
in a local Keplerian frame that follows a circular orbit in the ring
midplane. The actual value of $\vp$, which we associate with ``random
motion'' in this reference frame, depends on collision outcomes and
physical conditions in the disk, such as the number density of
particles $n(a)$, the surface (mass) density $\Sigma(a)$, and the
vertical scale height $h(a)$. These properties of the disk are
inter-related,
\begin{eqnarray}
n(a) & \sim & \frac{\Sigma}{2 h \mp}  
    \approx \frac{2\pi\Sigma}{\mp\vp\tkep}
\\
\label{eq:height}
h(a) & \sim & \frac{\vp \tkep}{4\pi},
\end{eqnarray}
where we assume a vertical speed $v_{\rm z} \approx \vp /
2$ \citep[e.g.,][and references therein]{oht1992}.  Thus, the microscopic
particle velocities and the broader structure of the disk are entwined
throughout its evolution.

\subsection{Particle collisions}\label{subsec:coll}

Physical collisions between particles hold the key to the evolution of
systems of satellites, moons, and planets. Collisions not only result in
fragmentation or mergers; they also damp random particle motions and
reduce collision speeds.  In the balance lies the difference between
growth or destruction of satellites in a ring. When particles collide,
their material properties determine the outcome. For icy bodies,
shattering or fragmentation occurs at speeds of $\vp \gae 10^3$~cm/s
\citep[][and references therein]{kb2004a}. At slower speeds, bouncing
and sticking can occur. Because high-speed collisions produce smaller
bodies that can dynamically cool larger ones, collisions are essential
to mergers and growth \citep[e.g.,][]{youdin2013}.

A first step toward assessing the role of collisions is to estimate
$\tcol$, the characteristic time between collisions for a single
particle. If particle velocities are comparable to or larger than 
their escape velocities, simple kinetic theory (``$nv\sigma$'') gives 
\citep[e.g.,][]{saf1969,liss1987,weth1993,gold2004}
\begin{equation}\label{eq:tcol}
\tcol \approx  \frac{1}{n\sqrt{2}\vp\pi 4\rp^2}
     \approx
\frac{\rhop\rp}{2\sqrt{2}\pi \Sigma} \tkep,
\end{equation}
where the cross-sectional area of $\pi (2\rp)^2$ is appropriate to
hard spheres, and the pairwise velocity $\sqrt{2}\vp$ is eliminated in
the rightmost expression because of its relationship to the scale
height, $h$ \citep[Equation~(\ref{eq:height})); see][]{oht1992}.  When
the relative speeds fall below the typical escape speed, gravitational
focusing increases the effective cross-section and reduces the
collision time
\citep[see \S\ref{subsubsec:gravrnd}, below]{weth1985a, weth1989, spaute1991,
gold2004, youdin2013}.

If collisions are inelastic, relative particle velocities are damped. 
Absent other effects, we can identify a damping time scale in terms of 
the evolution of random kinetic energy:
\begin{equation}
\frac{1}{\vp^2}\frac{d\vp^2}{dt} \sim -\frac{1}{\tdamp} ~ .
\end{equation}
Thus, the damping time is
\begin{eqnarray}\label{eq:tdamp}
   \tdamp & \approx & \frac{\tcol}{\floss}
\end{eqnarray}
where $\floss$ is the average fractional loss in kinetic energy per 
collision.  Since the random speed $\vp$ is measured in the local 
Keplerian frame, the damping time characterizes how collisions 
circularize the orbits of ring particles and flatten the ring.

The parameter $\floss$ carries the details of collisions between ring
particles.  Following \citet{porco2008}, we adopt simple 
parameterizations of the coefficients of restitution, which set
the ratio of the particle speeds before and after a collision.
Treating the normal (N) and transverse (T) directions separately, 
these coefficients are
\begin{eqnarray}
\epsilon_N & = & \min\left[\left(v_N/v_\ast\right)^{b},1\right]
\\
\epsilon_T & = & 0.9,
\end{eqnarray}
where $v_N$ is the normal component of the relative
velocity. Experiments suggest $v_\ast = 0.01$~cm/s and $b = -0.14$ 
for spheres of water ice \citep{sup1995,bridges1984,porco2008}.
Averaged over all impact parameters, the loss parameter for a given
collision speed $\vv$ is
\begin{eqnarray}
\flossv(\vv) & = & 1 - (\epsilon_N^2 + \epsilon_T^2)/2,
\end{eqnarray}
if collisions are random as in a 3D ideal gas. Averaging over the
distribution of relative speeds yields $\floss$ 
(Equation~(\ref{eq:tdamp})). In a simple model with
a Maxwell-Boltzmann distribution of relative speeds,
\begin{eqnarray}
\floss & \approx & \frac{\int_0^\infty \vv^4 d\vv \flossv(\vv) \exp(-3\vv^2/4\vp^2)}%
{\int_0^\infty \vv^4 d\vv \exp(-3\vv^2/4\vp^2)} 
  \\
    & \approx & 0.60 - 0.44 \left(\frac{\vp}{v_\ast} \right)^{-0.28} 
\ \ \ \ [b = -0.14, \vp \gg v_\ast].
\end{eqnarray}
In this case,
the energy loss for $\vp \gg v_\ast$ is limited by
grazing collisions and the value of $\epsilon_T$.  Unfortunately,
this value is not well constrained by experiment.

\subsection{Ring viscosity}\label{subsec:visc}

While physical collisions sap random kinetic energy from the ring, they
also add to it. By redirecting their trajectories after rebounding, 
collisions convert orbital motion into random velocities \citep{gold1978}.
Thus, the eccentricities and inclinations of ring particles grow; 
the entire ring broadens in semimajor axis. The diffusive spreading 
of ring particles can be expressed formally in terms of a viscosity,
\begin{equation}\label{eq:visc}
\nurad \sim 0.46 \frac{\vp^2 \tkep}{2\pi}\frac{\tau}{1+\tau^2} ,
\end{equation}
where the optical depth of particles in the ring is
\begin{equation}\label{eq:tau}
\tau \sim \frac{\pi \rp^2 \Sigma}{\mp} \sim \frac{3\Sigma}{4\rhop\rp}
\end{equation}
\citep{cook1964,gold1978,gold1982}. 
To illustrate how the viscosity $\nurad$ is associated with diffusive
spreading of the ring, we consider a particle as it rebounds during a
collision. If we are oblivious of the details, the collision outcome
appears stochastic, yielding a random step with a radial
component that is directed either inward or outward with equal
probability.  The magnitude of this step is roughly the particle's
epicyclic radius,
\begin{equation}\label{eq:rndstep}
|\delta a| \sim a \frac{\vp}{\vkep} .
\end{equation}
Viewing this outcome as part of a random walk, with steps taken at the
collision rate of $1/\tcol$ (Equation~(\ref{eq:tcol})), we can infer
an effective diffusion coefficient,
\begin{eqnarray}\label{eq:nuradest}
\nurad  \sim  \frac{|\delta a|^2}{\tcol} 
&  = & \frac{a^2\vp^2}{\vkep^2\tcol}
 =  \frac{a^2\vp^2}{(2\pi a/\tkep)^2\cdot(\rhop\rp\tkep/\Sigma)}
= \frac{\vp^2 \tkep\Sigma}{4\pi\rho\rp} 
=
\frac{\vp^2\tkep \tau}{3\pi},
\end{eqnarray}
where the rightmost expression is the low optical depth limit of
Equation~(\ref{eq:visc}) to within a factor of $\sim 2$. 
After $n$ steps (collisions), the typical radial displacement grows to 
\begin{equation}\label{eq:rndwalk}
|\delta a^{(n)}| \sim \sqrt{\nu t\,} \sim a \frac{\vp}{\vkep}
\sqrt{t/\tcol},
\end{equation}
with time $t$ set to $n\tcol$.  We can take advantage of
this formula in simulating particle diffusion. Using a random 
number generator we can send particles on random walks in 
orbital distance to mimic the behavior of a viscous ring
\citep[also \S\ref{subsec:rez}, below]{bk2013}.

Over many collisions involving an ensemble of particles,
the ring broadens by some radial distance $|\Delta a|$. 
The time scale for this spreading is
\begin{eqnarray}\label{eq:tspread}
\tspread & \sim  & \frac{|\Delta a^2|}{\nurad}.
\end{eqnarray}
This relationship explains the slow spreading of Saturn's rings.
Doubling the annular width of the A ring requires $\sim
1$~Gyr \citep{charnoz2009}. Slow spreading is a major challenge 
for models in which an impact ring around the Pluto-Charon binary 
expands in time to form the moons at their present location
\citep{kb2014}. We return to this issue in \S\ref{sec:pc}.

Because collisions move particles to regions with different Keplerian
velocity, they tend to pump up the relative speed of the particles. If
collisions are completely elastic, the relative velocities grow
unchecked \citep{gold1978}. For example, a particle in a sea of
objects on circular orbits might scatter slightly inward or outward
after a collision, so that it has a new speed relative to its new
neighbors, $|\Delta v| \sim e |\Delta a|$. Subsequent collisions 
cause that particle to walk randomly in velocity, as well as in radial
distance. By the $n^{\rm th}$ interaction the typical speed has grown
to
\begin{equation}\label{eq:vgrow}
|\Delta v^{(n)}|^2 \sim c |\Delta v^{(n-1)}|^2 \sim c^{n-1}|\Delta v|^2,
\end{equation}
where $c$ is a constant greater than unity that depends on the details
of the interactions (e.g., particle sizes, orbital distances).
Formally, random speeds quickly diverge \citep[e.g.,][]{gold1978}.

In reality, collisions involve some loss of kinetic energy,
providing a means to balance this process.  Coefficients of
restitution that decrease with speed --- and therefore cause more
kinetic energy loss at higher speeds --- allow a ring to reach an
equilibrium.  If the collisional damping is weak, rings 
equilibrate at high speeds; highly inelastic collisions lead to small
random motions and nearly circular orbits. The details of this balance
between dynamical heating and collisional damping are buried in the
magnitude of the equilibrium velocity, $\vp$ in Equation~(\ref{eq:visc}).

\subsection{Gravitational scattering}\label{subsec:grav}

In addition to physical collisions, pairwise gravitational encounters
can also cause random deflections and thus contribute to ring
viscosity \citep{horn1985,barge1990,oht1999,oht2002}. From
Equation~(\ref{eq:nuradest}), we can surmise how gravitational
interactions affect $\nurad$. Gravitational stirring and dynamical
friction \citep{weiden1989,kl1998,stewart:2000,kb2001} drive the local
velocity evolution; gravitational focusing makes the effective
cross-section larger and reduces the collision time. Thus, stirring
and focusing tend to increase the viscosity and drive the spreading of
the ring. However, for rings in a steady state, stirring pumps up
velocities until they exceed the escape speed of particles. Then,
gravitational focusing becomes unimportant (see below) as collisions
take over to regulate the velocities. Therefore, gravity's main role
in the ring viscosity is to help to maintain the equilibrium velocity
established by collisions. After all, only three things matter to the
spreading of the ring: how many particles are in the ring ($\Sigma$),
how big they are ($\rp$) and how fast they are moving relative to each
other ($\vp$).

Here we review some general aspects of gravitational interactions
between ring particles.  For particles outside the Roche limit,
separated by distances of a few Hill radii, we describe pairwise 
interactions as Rutherford scattering \citep{lin1979a}.
Working in a pair's center-of-mass frame, we relate impact parameters 
to scattering angles. Translating these scattering outcomes to changes 
in orbital elements, we then derive how gravity between ring particles 
affects the overall evolution of the ring.

\subsubsection{Rutherford scattering}\label{subsubsec:rutherford}

Pairwise gravitational interactions between massive particles are
described by Rutherford scattering. Following the traditional analysis
\citep[e.g.,][]{lin1979a}
we consider two point particles with masses $m_1$ and $m_2$ on
approaching trajectories in their center-of-mass frame with impact
parameter $b$ and closing speed $v$. Using the total and reduced masses
\begin{equation}
m = m_1+m_2 
\ \ \ \ {\rm and} \ \ \ \ 
\dm = \frac{m_1 m_2}{m_1 + m_2},
\end{equation}
we write equations of motion for the pair as a one-body problem in
terms of the pair's relative position and velocity. We are typically
interested only in the outcome of these encounters, described in terms
of
\begin{equation}\label{eq:scatang}
\Theta = 2 \arctan \frac{G m}{v^2 b},
\end{equation}
which is the angle between the initial and final relative velocities.
Another quantity that is important to
scattering outcomes is the distance of closest approach, $\rclose$,
derived from
\begin{equation}\label{eq:rclosesqr}
\rclose^2 = b^2 - \frac{2 G m \rclose}{v^2} =
 b^2 - \frac{\vesc^2}{v^2}(\rclose \cdot \rp),
\end{equation}
where $\vesc$ is the escape velocity of an individual particle of
radius $\rp$. Between the distance of closest approach and the
scattering angle, we distinguish between different outcomes: physical
collisions ($\rclose$ is less than the sum of particle radii) are one
possibility as described above. For purely gravitational encounters,
we also have strong/random scattering ($\Theta \sim 1$, $\rclose \ll
b$) and weak scattering ($\Theta \ll 1$, $\rclose \approx b$).  We
consider these regimes next, starting with strong interactions.

\subsubsection{Gravitational focusing and random 
scattering}\label{subsubsec:gravrnd}

When gravitational encounters are close and deflections are strong,
physical collision rates can be boosted and relative velocities can be
randomized. 
In a pairwise encounter, if the closing speed $v$ is small compared to
the escape velocity of the largest particle in a pair, then the
distance of closest approach $\rclose$ is significantly smaller than
the impact parameter $b$ (Equation~(\ref{eq:rclosesqr}). This effect is
gravitational focusing. If we compare the cross section for physical
collisions between identical particles to $b_c$, the maximum impact
parameter that leads to a collision, then we get the focusing factor,
\begin{equation}
\ffoc \equiv \frac{\pi b_c^2}{\pi (2\rp)^2} = 1+\frac{1}{2}\frac{v^2}{\vesc^2}.
\end{equation}
This factor quantifies the boost in the likelihood of physical
collisions and in the collision rate
\citep[in the context of planet formation, 
see][]{green1990,spaute1991,weth1993}. The overall rate of ring spreading 
is increased, with the viscosity parameter $\nurad$ augmented by a 
factor $\ffoc$.

Even in the absence of collisions, strong gravitational encounters can
affect the viscosity by randomizing velocities. When the impact
parameter $b$ is comparable to $Gm/v^2$, roughly the ``gravitational
radius'' of a particle, the scattering angle $\Theta$ covers the full
range of possibilities.  Thus, if the impact parameter is stochastic,
the scattered velocities are as well; close gravitational interactions
act like elastic collisions. Then, to derive the spreading of a ring,
we may adopt the formalism for collisions described in
\S\ref{subsec:visc}.  Thus, in the optically thin limit, there is an
effective viscosity that describes both physical collisions and
gravitational scattering. (For rigorous treatments of diffusive
spreading by gravity see \citealt{oht2003}, \citealt{ormel2012} and
\citealt{glaschke2014}.)

\subsubsection{Collective gravitational 
effects: migration}\label{subsubsec:migrate}

In addition to contributing to random motions, pairwise gravitational
interactions also act collectively to produce steady flows. Studies of 
planetary and satellite migration \citep{lin1979b,gold1980,ward1997} 
show how swarms of particles contribute over relatively long distances, 
$O(10) \rhill$ to cause a steady radial drift of mass in the ring.

To quantify this effect, we start with the Rutherford formula
(Equation~(\ref{eq:scatang})), and work only in the limit of small
scattering angle $\Theta$ since we are interested in larger-scale
collective effects, not small-scale 
random motions.  
As above, we consider two particles with
masses $m_1$ and $m_2$ on approaching trajectories in their
center-of-mass frame with impact parameter $b$ and closing speed $v$.
When the two particles are both orbiting a central body, scattering
through an angle $\Theta$ translates to changes in orbital elements
\citep{lin1979a,gold1980}. To see how this process plays out, we
first consider a pair of particles with extremely unequal masses.

We suppose $m_1 \rightarrow m$ and $m_2 \rightarrow \dm \ll m$ and
that the more massive particle is on a circular path around a central
object of mass $M$ at an orbital distance $a$.  Similarly, the lighter
body has an orbital distance of $a + \Da$, with $\Da/a\ll 1$, orbiting
on a circle in the same sense.  We can then write the denominator of
equation~(\ref{eq:scatang}) as a product of the pair's relative
orbital speed,
\begin{equation}
v \approx \Omega a \left(\frac{1}{\sqrt{1+\Da/a}-1}\right),
\end{equation}
and the relative orbital angular momentum
\begin{equation}
\ell \approx \dm \, \Omega a \, 
\Da \left(1 + \frac{\Da}{a}\right)  
\left[1-\frac{1}{(1+\Da/a)^{3/2}}\right],
\end{equation}
where $\Omega$ is the orbital angular velocity of the massive body. 
The change of angular momentum of the smaller object is
\begin{eqnarray}
\delta \ell & = & \left(1 -\cos\Theta\right)\ell \approx
\frac{1}{2}\Theta^2\ell,
\end{eqnarray}
in this small-angle, ``weak scattering'' limit.
Combining these results and expanding in terms of
$\Da$,
the change in orbital angular momentum is
\begin{equation}
\delta \ell \approx 
\frac{16}{9} \frac{G^2 m^2 \dm a}{\Omega^3 \Da^5}.
\end{equation}
This perturbation of the small satellite causes an equal and
opposite change in angular momentum of the larger particles, 
$\Delta\ell = -\delta\ell$.  The semimajor axis of the larger
object also changes.  For circular orbits,
\begin{equation}
\ell^2 \approx G M a.
\end{equation}
Taking derivatives of both sides of this expression gives the 
change in the larger body's semimajor axis as
\begin{equation}\label{eq:torquelike}
\delta a  \approx -\frac{2 \delta \ell}{m\Omega a}
\end{equation}
Applying Kepler's third law ($GM = \Omega^2 a^3$):
\begin{equation}\label{eq:daxtreme}
\delta a  \approx  
- \frac{32}{9} \frac{m \, \dm}{M^2} \frac{a^6}{\Da^5}
\left (1 + \frac{9}{4} \frac{\Da}{a} + ... \right)
\ \ \ \ \ \ \ \ [m \gg \dm]
\end{equation}
A similar analysis of scattering between equal mass objects
gives
\begin{equation}
\label{eq:daequal}
\delta a \approx - \frac{256}{25} \frac{m^2}{M^2 }\frac{a^6}{\Da^5}
\left (1 + \frac{27}{20} \frac{\Da}{a} + ... \right)
\ \ \ \ \ \ \ \ \ \mbox{\rm [equal-mass satellites]}
\end{equation}
The change in orbital distance per encounter per unit
mass is 
\begin{equation}\label{eq:dadm}
\dadm(a,\Da) \approx - \frac{8 A m}{M^2}\frac{a^6}{\Da^5} 
\left (1 + B \frac{\Da}{a}\right),
\end{equation}
where $A = 4/9$ and $B = 9/4$ when $m\gg\dm$, and $A = 32/25$
and $B = 27/20$ 
when the mass of the scatterers is equal.
Thus, in general $A = 0.4$--1.3 and $B = 1.3$--2.3 (the 
factor of 8 in Equation~(\ref{eq:dadm}) was chosen to keep these
constants near unity).

Equation (\ref{eq:dadm}) is valid only for a limited range of orbital
configurations.  The Hill radius of the larger satellite, which
characterizes that particle's domain of influence, provides a lower 
limit to the orbital separation $\Da$.  Furthermore, for the weak theory 
to apply, an interaction with separation $\Da$ must yield small orbital 
perturbations. This condition is violated if the orbital separation is 
within
\begin{equation}\label{eq:rxing}
\rxing \equiv 2\sqrt{3}\rhill \approx 3.5\rhill
\end{equation}
since the outcome of such encounters is that the pair chaotically 
cross orbits\footnote{For exactly circular orbits in the extreme 
mass-ratio case, weak deflections can cause orbits to cross when 
the orbital separation is less than $2\sqrt{3}\rhill$.}
\citep{glad1993}.
The theory breaks down at large orbital separations, where the 
curvature of the satellites' orbits plays a role in the interactions 
\citep{bk2011b}. Combining these two constraints gives
\begin{equation}\label{eq:dalimits}
     \rxing \lae \Da \lae 0.1 a
\label{eq:limits} 
\end{equation}
which is an approximate condition for Rutherford-style weak
scattering.

Migration theory uses these conclusions to predict the drift rate of a
satellite in a particle disk by tracking the rate of encounters along
with the orbital displacement per encounter \citep[e.g.,][]{ida2000b}.
This same strategy applies to a ring of equal-mass particles. The
interaction rate between pairs of particles at orbital separation
$\Da$ is set by their synodic period:
\begin{eqnarray}\label{eq:Psynodic}
\tsynodic(a,\Da) & \approx & \frac{2 a \tkep}{3|\Da| \left[1 -
    5\Da/(4a)\right]} \ ,
\end{eqnarray}
Integration over the disk gives the total drift
rate,
\begin{eqnarray}\label{eq:dadt}
\dadtweakgrav & \approx & 
\int 2\pi (a+q) d\Da \hipass(\Da) 
     \frac{\Sigma(a+\Da)\dadm(a,\Da)}{\tsynodic(a,\Da)},
\end{eqnarray}
where $\hipass(q)$ is a high-pass filter, set to zero at small scales
(e.g., $|q| \lae 3\rhill$) and unity elsewhere.  Thus the filter eliminates
contributions to the drift rate on scales where the weak-scattering
theory breaks down and where interactions are dominated by random
motions.

With the weak-scattering result for $\delta a$,
and a Taylor expansion in terms of the radial distance variable $\Da$,
the drift rate becomes
\begin{eqnarray}\label{eq:dadtx} 
\dadtweakgrav & \approx & -\frac{24 A \pi a^2 \Sigma}{M} \frac{m}{M}
\frac{a}{T} \int \mbox{\rm sgn}(\Da) \hipass(\Da) \frac{d\Da}{a}
\frac{a^4}{\Da^4} \times
\left[1 + \frac{\Da}{a}
\left({B}-\frac{1}{4}+\frac{a}{\Sigma}\frac{d\Sigma}{da}\right)\right],
\end{eqnarray}
where the constants $A$ and $B$ are of order unity and depend on
the mass ratio of the drifting particle and other objects in the disk
(see Equations~(\ref{eq:daxtreme}) and (\ref{eq:daequal})).

In a dynamically cold ring --- where the scale height $h$ is smaller
than the Hill radius of the ring particles --- the minimum orbital
distance between particles for which steady migration is possible is
$\Damin \approx \rxing$ (Equation~(\ref{eq:rxing})). Inside this
distance, orbits are chaotic and can cross, enabling physical collisions. 
Outside this scale, orbital evolution is diffusive. The ring of particles
expands until particle separations become very large 
(e.g., Equation~(\ref{eq:dalimits})). 

If the ring is dynamically hot, with random particle excursions 
comparable to the scale height $h$, then $\Damin \approx h$.  By setting 
$\hipass(\Da)$ = 0 when $|\Da|<\Damin$ and unity elsewhere, we obtain a
general expression for the radial migration of a ring particle: The radial
drift rate of a particle embedded in the ring is
\begin{eqnarray}
\label{eq:dadtweakgrav}
\dadtweakgrav & \approx & 
-\frac{24 \pi a^2 \Sigma}{M}
\frac{m}{M}\frac{a}{T} 
\frac{a^2}{\Damin^2}
A
\times 
\left({B}-\frac{1}{4}+\frac{a}{\Sigma}\frac{d\Sigma}{da}\right)
\end{eqnarray}
where 
\begin{equation}\label{eq:damin}
\Damin = \max(\rxing,h) \ \ \ \textrm{or} \ \ 
\ \Damin^2 \approx (16\rhill^2 + h^2).
\end{equation}
The rightmost expression just serves to enforce the minimum
value of $\Damin$ as a smooth function of $h$ or $\rhill$.
In the expression for the drift rate itself, the quantity $B-1/4$
represents a systematic inflow of geometrical origin in a disk with
uniform surface density, while the logarithmic derivative corresponds
to a radial flux in a non-uniform disk analogous to diffusion in a
viscous ring.  Together these two terms define an equilibrium surface
density with no radial flux ($d\ln\Sigma/d\ln a = 1/4-B$).  Near the
inner edge of the ring, where the radial gradient of $\Sigma$ is
large, material flows inward. Similarly, the ring spreads outward near
the outer edge.

\subsection{Ring evolution}

Astrophysical rings continuously evolve with time. Physical collisions
and gravitational interactions produce diffusive spreading, growth,
and destruction. Inside the Roche limit, simplifying assumptions, such
as requiring fixed particle size \citep{gold1978,salmon2010}, allow
elegant solutions to this evolution.  Here, our goal is to track
simultaneous growth and viscous spreading. We take advantage of the
idea that pairwise interactions drive all the dynamics. In a steady
ring, most interactions involve scattering and are neither destructive
or fully inelastic (all collisions lead to perfect mergers).  Then, the 
local dynamics (viscous or gravitational spreading) can be handled by
approximating particles as having a fixed size over a certain short
time scale to track the dynamical evolution of the ring.  Particle
growth can then enter as a separate step in the evolution, in staggered 
fashion, in the spirit of a standard leap-frog technique.
This approach also allows us to use standard tools of the trade
\citep{pri1981} to calculate the ring evolution.

\subsubsection{Steady-state theory of ring evolution}

We describe the evolution of a ring in terms of its surface density.
The evolution equation follows from conservation of mass
and angular momentum \citep{pri1981}.  Integrated over the direction
perpendicular to the ring plane, these conservation laws are
\begin{equation}\label{eq:continuity}
\frac{\partial \Sigma}{\partial t} +
\frac{1}{a}\frac{\partial}{\partial a}\left(a\dot{a}\Sigma\right)
=0,
\end{equation}
\begin{equation}\label{eq:angmoconserve}
\frac{\partial}{\partial t}
\left(a^2\Omega\Sigma\right)
+ \frac{1}{a} \frac{\partial}{\partial a}\left( a^3\dot{a}\Omega\Sigma \right) 
= \frac{1}{2\pi a}\frac{\partial \dot{L}}{\partial a},
\end{equation}
where $a$ is orbital distance, $\Omega$ is orbital angular speed at
distance $a$, and $\dot{L}$ represents the torque on an annulus of 
radius $a$ and width $\Delta a$.  These equations come from applying
the total time derivative operator $d/dt$ in cylindrical coordinates
to the total mass and angular momentum associated with the annulus. In
the case of angular momentum conservation, the derivative of the
torque $\dot{L}$ follows from taking the limit of $\Delta a
\rightarrow 0$.  Combining these expressions we obtain
\begin{eqnarray}
a \dot{a}\Sigma & = & 
  \left[\frac{\partial}{\partial a}\left(a^2\Omega\right)\right]^{-1}
\frac{1}{2\pi} \frac{\partial\dot{L}}{\partial a}
 \approx \frac{1}{\pi a\Omega} \frac{\partial\dot{L}}{\partial a}
\end{eqnarray}
where the expression on the right applies in the limit of Keplerian flow.
Putting this expression back into the continuity equation gives
the evolution equation for the surface density:
\begin{equation}\label{eq:ringevolveraw}
\frac{\partial\Sigma}{\partial t} 
=
-\frac{1}{a}\frac{\partial}{\partial a}\left(a\dot{a}\Sigma\right)
=
-\frac{1}{a}\frac{\partial}{\partial a}\left(\frac{1}{\pi a \Omega}
   \frac{\partial\dot{L}}{\partial a}\right) .
\end{equation}
It remains to quantify the torque, either in terms of $\dot{L}$ or the
corresponding radial drift $\dot{a}$. Multiple sources of torque are
just summed together, using both $\dot{L}$ (for viscosity) and
$\dot{a}$ (for gravitational spreading).  The latter we include in the
form of Equation~(\ref{eq:dadtx}), while the former we calculate next,
following \citet{pri1981}.

The viscosity in an astrophysical ring relates the shear in the
velocity flow ($a\partial\Omega/\partial a$, which is $-3/2\Omega$
for a Keplerian ring) to the torque on an annulus,
\begin{eqnarray}\label{eq:visctorque}
\dot{L} 
 & = & \nu a \cdot 2\pi a \Sigma \cdot a \frac{\delta\Omega}{\delta a}
\approx  -3\pi a^2 \Sigma \nu \Omega.
\end{eqnarray}
These expressions define the kinematical viscosity $\nu$. They may be
easily related to the laboratory definition of dynamical viscosity by
specifying a disk scale height and working with the force
applied to the common surfaces between an annulus and its neighbors.

Putting the above result (Equation~(\ref{eq:visctorque})) together
with the expression for the radial drift rate from weak scattering
(Equation~(\ref{eq:dadtweakgrav})) yields the desired evolution
equation:
\begin{equation}\label{eq:ringevolve}
\frac{\partial\Sigma}{\partial t} 
=
\frac{1}{a}\frac{\partial}{\partial a}
\left[3 a^{1/2} \frac{\partial}{\partial a}
\left(\nu \Sigma a^{1/2}\right)
- \left(a\Sigma\dadtweakgrav\right)
\right].
\end{equation}
In this expression, the first term on the right-hand-side gives the
diffusive evolution of $\Sigma$ from viscosity $\nu$.  The second term
corresponds to gravitational scattering.  Under some circumstances,
for example, a ring of monodisperse particles with a shallow surface
density gradient and relative speeds higher than the escape velocity,
we can derive an approximate viscosity for the gravitational
interactions,
\begin{equation}
\nu_{\rm grav} \approx 
\frac{\vesc^2 \Sigma\tkep}{\rho r} 
\frac{\vesc^2}{v^2}  
\ \ \ \ \ \  (v > \vhill).
\end{equation}
This expression is comparable, to within a constant of order unity, to
similar expressions derived from the standpoint of kinetic theory
\citep{shu1985}. Since we derive the ring evolution directly from
scattering theory, and do not seek a quantity of the form of a
viscosity parameter, our theory differs slightly from the kinetic
approach.  In principle, an advantage of our approach is that it can
accommodate migration when the distribution of ring particle mass
covers a wide range.  Below (\S\ref{subsubsec:viscgrav}) we further
discuss the connection between the evolution from the effective
(collisional) viscosity and the contribution from gravity in our
formalism.

The solution of Equation~(\ref{eq:ringevolve}) requires boundary
conditions.  In general, the outer edge of the ring might expand into
a vacuum (or be truncated artificially at the edge of a computational
domain). The inner boundary is more sensitive to the astrophysical
setting.  With a single central mass, material may be lost by
accretion, thus $\Sigma = 0$ at its surface.  For a circumbinary ring,
the lack of stable orbits within a factor of at least twice the binary
separation (for comparable-mass binaries) establishes the inner
boundary of the disk. Preliminary simulations show that particles at
the inner edge get pushed back into the ring, or are trapped in
resonances there (\S\ref{subsec:rez}, below). We therefore
provisionally set inner boundary condition at the innermost stable
circular orbit to be reflective, $d{\Sigma}/da = 0$ \citep[see
  also][]{pri1991}.

Together with specific values for $\nu$ (\S\ref{subsec:visc}) and the
drift rate from mutual weak scattering of ring particles
(\S\ref{subsec:grav}), these boundary conditions allow us to track
the spreading of a ring from some initial state.

\subsubsection{Connection between weak scattering and ring viscosity}
\label{subsubsec:viscgrav}

Ring viscosity and gravitational scattering have similar physical
origins. In both cases, interactions between pairs of particles cause
deflections which lead to a radial mass flux. There are also profound 
differences.  Deflections in a viscous ring derive from physical collisions,
which can stir particles up until collisional damping becomes
important \citep{cook1964,gold1978}.  Gravitational scattering can
also stir particles, but only until they reach the escape speed. Beyond 
that speed, gravitational scattering becomes ineffective. Furthermore, 
gravitational interactions become irrelevant if the particles lie inside 
the Roche limit.  In both of these situations, the ring is completely 
driven by the viscosity from physical collisions.  Otherwise, gravity 
and viscosity share in the evolution of the ring.

To compare quantitatively how viscosity and gravity operate,
we focus on the surface density flux terms (right hand side of
Equation~(\ref{eq:ringevolve})).  Under a set of simplifying
assumptions, including constant viscosity $\nu$ and surface density
$\Sigma$ in a ring of monodisperse particles, we write the fluxes
in terms of the particle escape velocities $\vesc$:
\begin{eqnarray}
\label{eq:fluxvisc}
\fluxvisc & \equiv & \frac{3}{2}\nu\Sigma 
\sim G\Sigma^2 \rp \tkep \frac{\vp^2}{\vesc^2}
\\
\label{eq:fluxgrav}
 \fluxgrav & \equiv & a \Sigma \dadtweakgrav
\sim G\Sigma^2 \rp \tkep \frac{\vesc^2}{\vp^2 + C^2 \vhill^2} 
\end{eqnarray}
where $C \sim \rxing/\rhill \sim 4$.  Thus, the particle velocity, as
compared to the escape velocity, delineates the viscous and gravity
regimes. Roughly, viscosity operates at $\vp \gg \vesc$, while
gravitational scattering drives the evolution at $\vp \ll \vesc$.

In a viscous disk, the effective viscosity formally depends on the random
velocity with no restrictions except that $\vp$ must small compared to
the local Keplerian velocity. In the gravity dominated regime, 
the surface density flux has a maximum,
\begin{equation}
 \fluxgrav 
\sim G\Sigma^2 \rp \tkep \frac{\vesc^2}{16\vhill^2}  \ \ \ (\vp \ll \vesc).
\end{equation}
Physically this limit reflects our assumptions about gravitational
interactions: particles with low relative speeds drifting by one
another with impact parameters within $\rxing$ are in the chaotic,
orbit crossing regime, or possibly corotating. Here, we assume that
these types of orbits do not lead to a net radial drift.  In any
event, if the speeds are below the escape velocity, then collisional
damping also must be operating, keeping the ring dynamically cool on
these small scales.

\subsection{Illustrations of the theory}

To give examples of solutions to the evolution equation
(\ref{eq:ringevolve}), we use a finite difference code that divides
the surface density into 640 radial bins. (In runs described below, we
doubled or quadrupled the number of bins to track the physics of any
sharp edges that develop.) The initial surface density has a Gaussian
profile in orbital distance, and a fixed total mass. In each run, we
choose a particle radius and set the dispersion of the particles to be
their escape velocity. These parameters, along with the surface
density on the grid and its gradient (derived from finite differences)
give the viscosity $\nurad$ and $\dadtweakgrav$ needed to evolve
$\Sigma$ in a step-wise fashion.

\begin{figure}[htb]
\centerline{\includegraphics[width=7.0in]{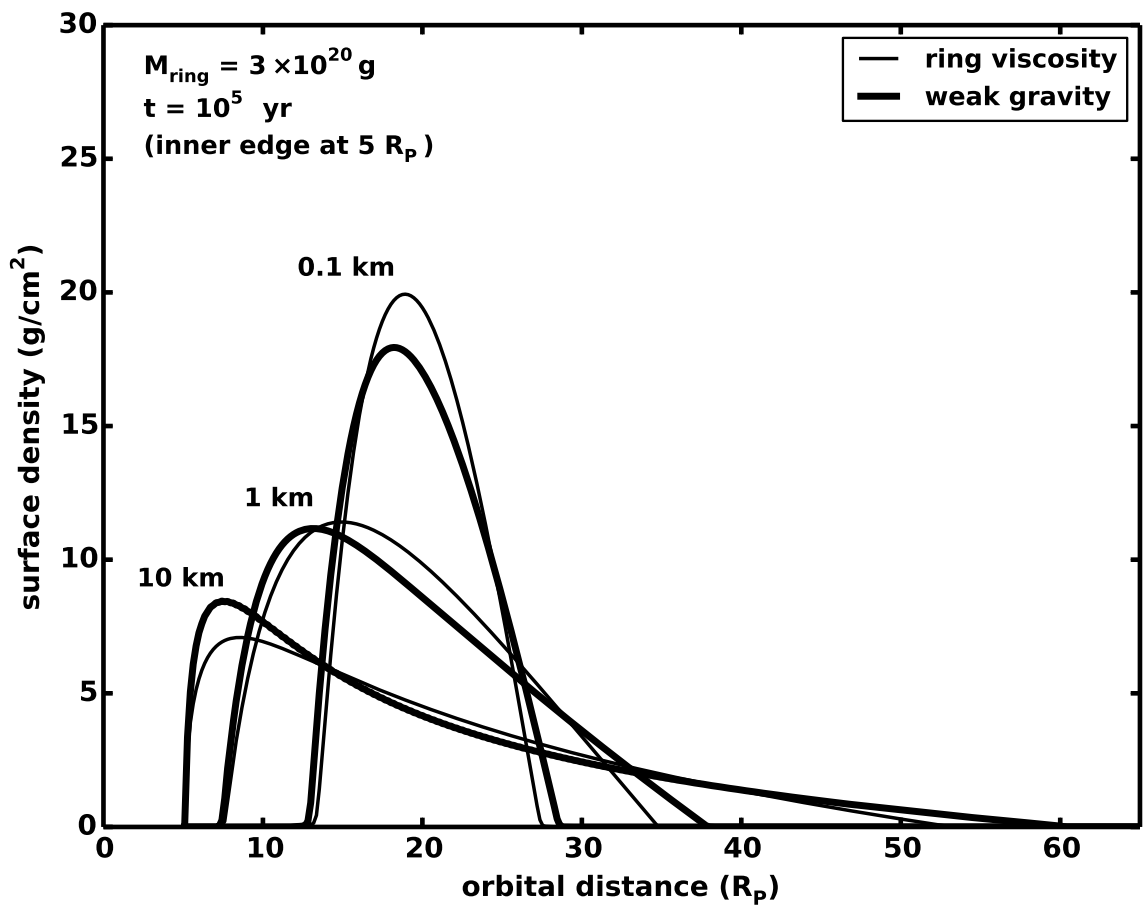}}
\caption{\label{fig:spreadPCgrav-v-visc} A comparison between the
  effects of viscosity and gravitational scattering in the spreading
  of rings. The astrophysical context is Pluto-Charon, at some early
  time when the binary is compact (orbital separation $\aPC = 5\,\RP$)
  and surrounded by a $3\times 10^{20}\,$g ring.  Each scenario 
  begins with the same surface density profile in the ring, but
  differ in the size of the ring particles, as indicated.  In each
  case, the particles are stirred to their escape velocity. The dark 
  curves spread from gravity only; light curves spread solely from 
  viscosity. At this choice of particle speed, the two mechanisms 
  should be similar. At higher speeds viscosity will dominate; 
  at lower speeds gravity will drive the spreading.  
}
\end{figure}

Figure~\ref{fig:spreadPCgrav-v-visc} show numerical solutions for the
evolution of a ring around a central mass.  The curves in the Figure
are snapshots of the evolution, corresponding to a viscous ring and to
an inviscid ring that spreads from gravity.  The two phenomena yield
similar behavior in the ring at this delimiting velocity.

For reference, we consider how the theory applies to Saturn's
rings.  Inside the Roche limit, mutual gravity between small ring 
particles is irrelevant. Outside this limit, external influences 
(distant moons) stir the ring particles to speeds that exceed their 
escape velocity.  For example, the A ring consists of small 
(centimeter- to meter-size) particles, with a scale height of roughly 
10 meters \citep[e.g.,][]{colwell2009}. The orbital frequency is 
$\Omega \sim 10^{-4}$~s$^{-1}$; typical random speeds, $\vp \sim h\Omega$, 
are $\sim$0.1~cm/s. The escape velocity of a centimeter-size particle is
roughly $0.001$~cm/s; thus, viscosity dominates in the ring.

In other planetary rings, conditions probably differ.  Our focus turns
to the Pluto-Charon binary where particle masses may have been large
enough for gravity to have an impact on ring evolution.

\section{Circumbinary orbital dynamics}\label{sec:circumbin}

Pluto and Charon, as partners in the solar system's most prominent
binary planet, pose an interesting challenge to our understanding of
the dynamics of planetary rings and satellites.  Circumbinary orbits
differ from Keplerian ones in important ways
\citep{hepp1978,murray1999,lee2006,leung2013,bk2015tatooine}.  In this
section, we present an overview of these differences.

\subsection{Orbital stability and resonances}\label{subsec:rez}

At distances comparable to the binary separation, $\abin$,
circumbinary orbits are unstable. To help map out regions of
instability around a binary, \citet{holman1999} numerically estimate a
critical distance from the binary center of mass, $\acrit$, that
depends on the binary eccentricity ($\ebin$) and mass ratio. Beyond
$\acrit$, most orbits are stable over many ($10^4$) dynamical times. 
Inside $\acrit$, the gravity of the binary destabilizes orbits.  Thus,
every binary has an inner cavity that is cleared of ring particles.  
For Pluto-Charon in its present-day orbit ($\ebin \approx 0$), this 
inner cavity lies inside of $\acrit \approx 2.0 \abin$.  With
$a \sim 2.2\abin$, the innermost moon Styx is just beyond this 
critical orbital distance.

Instabilities arise
when resonances with the central binary overlap
\citep{wisdom1980,lecar2001}. The \citet{holman1999} expression for
$\acrit$ gives only an approximate location of the outer edge of the
central cavity around the binary.  Some stable orbits exist inside of
$\acrit$; there may be various unstable resonances outside of it.
\citet{popova2013} provide an excellent illustration
\citep[seea also][]{wyatt1999, pichardo2008, doolin2011}.  

For the Pluto-Charon binary, with eccentricity near zero, resonance
phenomena are not important except near the 3:1 resonance at $\sim
2\abin$. When the binary has a modest eccentricity, $\ebin = 0.1$,
however, unstable orbits form around the 5:1 commensurability.  For
other values of $\ebin$, it is important to map the stability orbits
for each astrophysical setting, especially for problems which involve
massive satellites \citep[e.g.][]{popova2013,cheng2014a}.

Figure~\ref{fig:mostcircpcx} illustrates the effects of resonances. 
Initially, a massless satellite orbits outside of the 7:1 commensurability 
around an expanding circular binary. On its own, the satellite navigates
the 7:1--4:1 commensurabilities as binary expansion causes them to
sweep past. Near the 3:1 resonance (with $a$ close to $\acrit$), however, 
the satellite is ejected (Fig.~\ref{fig:mostcircpcx}, left panel).

In contrast, a satellite that experiences sufficient collisional damping 
and diffusion survives. In Fig.~\ref{fig:mostcircpcx} (right panel), a
satellite starts in the same orbit but is subjected to diffusion and
random walks in a manner consistent with a monodisperse ring of 1-km
particles stirred to their escape velocity in a ring with surface
density 40~g/cm$^2$ (see \S\ref{subsec:pcinit}, below).  We mimic
these effects in the manner of \citet[e.g.,
Equation~(\ref{eq:rndwalk}) and related discussion, above]{bk2013};
damping comes from a slow and steady adjustment of the osculating
Keplerian eccentricity, while diffusion is a sequence of random walks.

With collisional damping and diffusion, satellites in the rings
coexist with the destabilizing resonances. Strong damping (many small
particles) allows particles to orbit as if the instability did not
exist; weak damping (fewer, larger particles) and higher binary
eccentricity leaves satellites susceptible to ejection.  Intermediate
levels of damping and diffusion allow particles to become trapped in
resonances and to move with them.  In the example shown in the left
panel of Fig.~\ref{fig:mostcircpcx}, stable satellites expanding in
the 3:1 commensurability require damping times $\lesssim$ 50~yr
(more than $10^3$ satellite orbital periods). When all of the mass
in the ring is in objects with radii smaller (larger) than a few km,
the time scale for collisional damping (cf. Equation~(\ref{eq:tdamp})
is smaller (larger) than this requirement. Thus, different mixtures
of particle sizes lead to different outcomes for the evolution of 
the Pluto-Charon binary and its satellites.

\subsection{Circumbinary orbits and most circular paths}

Outside the resonances, circumbinary orbits are stable and can be
surprisingly circular. Despite the non-Keplerian gravitational
potential of the binary, there is a set of orbits analogous to
circular paths around a single central mass.  This family of `most
circular' orbits consists of nested, circumbinary trajectories that do
not cross \citep{lee2006,youdin2012,leung2013}.  Even when the orbit
of the central binary is eccentric, the most circular paths have
eccentric motion characterized by a `forced' eccentricity, $\eforce$
\citep{hepp1978,murray1999} that preserves the nested and
non-intersecting properties.

Particles with random motion about a most circular orbit have a
`free' eccentricity, $\efree$, which serves the same role as the orbital
eccentricity for an orbit around a single central mass. Particles with
free eccentricity collide with characteristic relative velocities
$\efree\vkep$.  Particles with eccentricity that damps to zero settle 
on most circular paths with a forced eccentricity.

\subsubsection{Analytical theory of circumbinary orbits}

Most circular orbits appear as part of general circumbinary orbit
solutions from analytical theory in the context of the restricted 
three body problem. Following \citet{lee2006} and \citet{leung2013}, 
we start with the potential
\begin{equation}\label{eq:Phi}
\Phi = -\frac{G \Mp}{\sqrt{R^2+z^2+\Rp^2+2R \Rp \cos\Delta\phi}}
- \frac{G \Ms}{\sqrt{R^2+z^2+\Rs^2-2R \Rs \cos\Delta\phi}},
\end{equation} 
where $\Delta\phi$ is the angle between the secondary and the massless 
satellite (ring particle) in a reference frame with the binary's
center of mass at the origin; in this frame, the satellite is at
radial position $R$ in the plane of the binary, and altitude $z$ above
it, while $\Rp$ and $\Rs$ are the orbital distances of the primary and
secondary, with masses $\Mp$ and $\Ms$. When the binary is
circular ($\ebin = 0$), $\Rp$ and $\Rs$ are constants.

The strategy of \citet{lee2006} is to expand the potential in terms of
the angle cosines in Equation~(\ref{eq:Phi}): for a circular binary,
this expansion is
\begin{eqnarray}\label{eq:phiseries}
\Phi & \approx & \sum_{k = 0}^{\infty}
\Phi_{0k}\, \cos(k\Delta\phi) 
\ \ \ \ \ \ \ \ \ \ \ [\ebin = 0;\ z = 0]
\end{eqnarray}
where $\nbin$ is the mean motion of the binary ($\nbin^2 =
G(\Mp+\Ms)/\abin^3$), and our choice of time $t$ fixes the orbital
phase. Each coefficient $\Phi_{0k}$ may be expressed as a series in
terms of $\rgc/\abin$, where $\rgc$ is the radius of a guiding center
associated with a satellite's orbit.  \citet{lee2006} and
\citet{bk2015tatooine} list leading terms in these expansions.

The next step is to express a satellite's motion as excursions in
radial, azimuthal and altitude coordinates ($\delta R$,$\delta
\phi$,$z$) about the guiding center at radius $\rgc$. The result is a
set of equations of motion that can be linearized in the excursion
coordinates. When set up in the manner of driven harmonic oscillators,
these equations can be solved with standard techniques. The natural
frequencies $\kappae$ and $\kappai$ correspond to the free
eccentricity and free inclination. The driving frequencies are the
synodic frequency and its harmonics, along with the binary's orbital
frequency if the binary is eccentric.

This prescription yields expressions for the satellite position in
cylindrical coordinates \citep{lee2006}:
\begin{equation}
\label{eq:mostcircr}
  R(t) = \rgc\left[1 - \efree\cos(\kappae t+\psi_e) - \eforce\cos(\ngc t) +
     \sum_{k=1}^\infty {C_k}\cos(k\nsyn t)+  \ldots\right] 
\end{equation}
\begin{equation}
\label{eq:mostcircphi}
  \phi(t) = \ngc\left[ t + \frac{2\efree}{\kappae} 
\sin(\kappae t+\psi_e) + \eforce \sin(\ngc t)
+ 
\sum_{k=1}^\infty \frac{D_k}{k\nsyn}\sin(k\nsyn t)+ \ldots\right] 
\end{equation}
\vspace*{5.5pt}
\begin{equation}
\label{eq:mostcircz}
z(t) = i \rgc \cos(\kappai t + \psi_i),
\end{equation} 
where $\ngc$ is the angular speed of the guiding center, $\nsyn \equiv
\nbin-\ngc$ is the synodic frequency of the satellite relative to the
binary; orbital elements are $\eforce$, $\efree$ and $i$, the forced
eccentricity, the free eccentricity and the inclination, respectively,
while the phase angles $\psi_e$ and $\psi_i$ are constants.  The
coefficients $C_k$ and $D_k$ are associated with the time-varying part
of the binary's potential. Terms not shown depend on the binary
eccentricity \citep[see][]{leung2013}. In the form given above, the
solutions are accurate when $\ebin \lae \rgc/\abin \ll 1$.

To define the parameters in
Equations~(\ref{eq:mostcircr})--(\ref{eq:mostcircz}), we start by
noting that the orbital frequency of the guiding center comes from the
time-average potential $\Phi_{00}$ (Equation~(\ref{eq:phiseries})):
\begin{equation}
\ngc^2  \equiv \frac{1}{\rgc} \left.\frac{d\Phi_{00}}{dR}\right|_\rgc
= 
  \frac{G M}{\rgc^3}\left\{1 + \frac{\mreduced}{M}\left[
    \frac{3}{4}\frac{\abin^2}{\rgc^2}  
     + \frac{45}{64}\frac{(\Mp^3+\Ms^3)}{M^3}\frac{\abin^4}{\rgc^4}
     + ... \right]\right\},
\end{equation}
the square root of which is the mean motion of the ring particle.
Its epicyclic frequency is 
\begin{equation}
\kappae^2 \equiv \rgc \left.\frac{d\ngc^2}{dR}\right|_\rgc\!\!-4\ngc^2
 = 
  \frac{G M}{\rgc^3}\left\{1 - \frac{\mreduced}{M}\left[
     \frac{3}{4}\frac{\abin^2}{\rgc^2}  
     + \frac{135}{64}\frac{(\Mp^3+\Ms^3)}{M^3}\frac{\abin^4}{\rgc^4}
     + ...\right]\right\}.
\end{equation}
A similar analysis based on the expansion of the potential in the
$z$-direction yields the vertical excursion frequency,
\begin{equation}
\kappai^2 \equiv \left. \frac{1}{z}\frac{d\Phi}{dz}\right|_{z=0,\rgc}
= \frac{G M}{\rgc^3}\left\{1 + \frac{\mreduced}{M}\left[
     \frac{9}{4}\frac{\abin^2}{\rgc^2}
     + \frac{225}{64}\frac{(\Mp^3+\Ms^3)}{M^3}\frac{\abin^4}{\rgc^4}
     + ...\right]\right\}.
\end{equation}
Because $\kappae$ and $\kappai$ differ from the mean motion of the
satellite, the free eccentricity and inclination have corresponding
nodal precession.  In the limit that the binary separation goes to zero,
both $\kappae$ and $\kappai$ become the Keplerian mean motion.  Next,
the coefficients for the forcing terms at the synodic frequency are
\begin{eqnarray}
  C_k & = & 
\frac{1}{\rgc(\kappae^2-k^2\nsyn^2)}
\left[-\frac{d\Phi_{0k}}{dR}
  +\frac{2\ngc\Phi_{0k}}{\rgc\nsyn}\right]_{\rgc}
\\
  D_k & = & 2 C_k - \frac{\Phi_{0k}}{\rgc^2\ngc\nsyn},
\end{eqnarray}
in the notation of \citet{lee2006}.
Finally, we have the forced eccentricity,
\begin{equation}\label{eq:eforce}
\eforce =  \frac{5}{4}\frac{\Mp-\Ms}{M} \frac{\abin}{\rgc} \ebin 
- \left[\frac{1}{2}\frac{\mu(\Mp-\Ms)}{M^2} + 
  \frac{5}{32}\frac{\Mp^3-\Ms^3}{M^3}\right]\frac{\abin^3}{r^3}
+ ...
\end{equation}
\citep{hepp1978,mori2004,raf2013}. \citet{leung2013} include this term
($C^-_1$ in their notation), along with terms similar to $C_k$ and
$D_k$ but reduced by a factor of the binary eccentricity. 

\subsubsection{Examples}

Most circular paths, with no inclination or free eccentricity,
describe the least dissipative trajectories for particles that suffer
collisions.  Small particles on these paths remain on them
indefinitely without colliding.  We construct these trajectories
using the orbit solutions in
Equations~(\ref{eq:mostcircr})--(\ref{eq:mostcircz}).  Because the
fall-off with $\abin/\rgc$ is shallow for orbits that are modestly
close to the binary, we also consider numerical search methods. We 
evolve many trial paths until we find the one that minimizes the
radial excursion \citep[e.g.,][]{youdin2012}.  This procedure is 
slow, as it requires numerical orbit integration with each trial.  For
the integrations, we use a 6$^{\rm th}$-order symplectic time step
algorithm in the N-body component of our \textit{Orchestra} code
\citep{bk2006,bk2011a}.

\begin{figure}[htb]
\centerline{\includegraphics[width=7.0in]{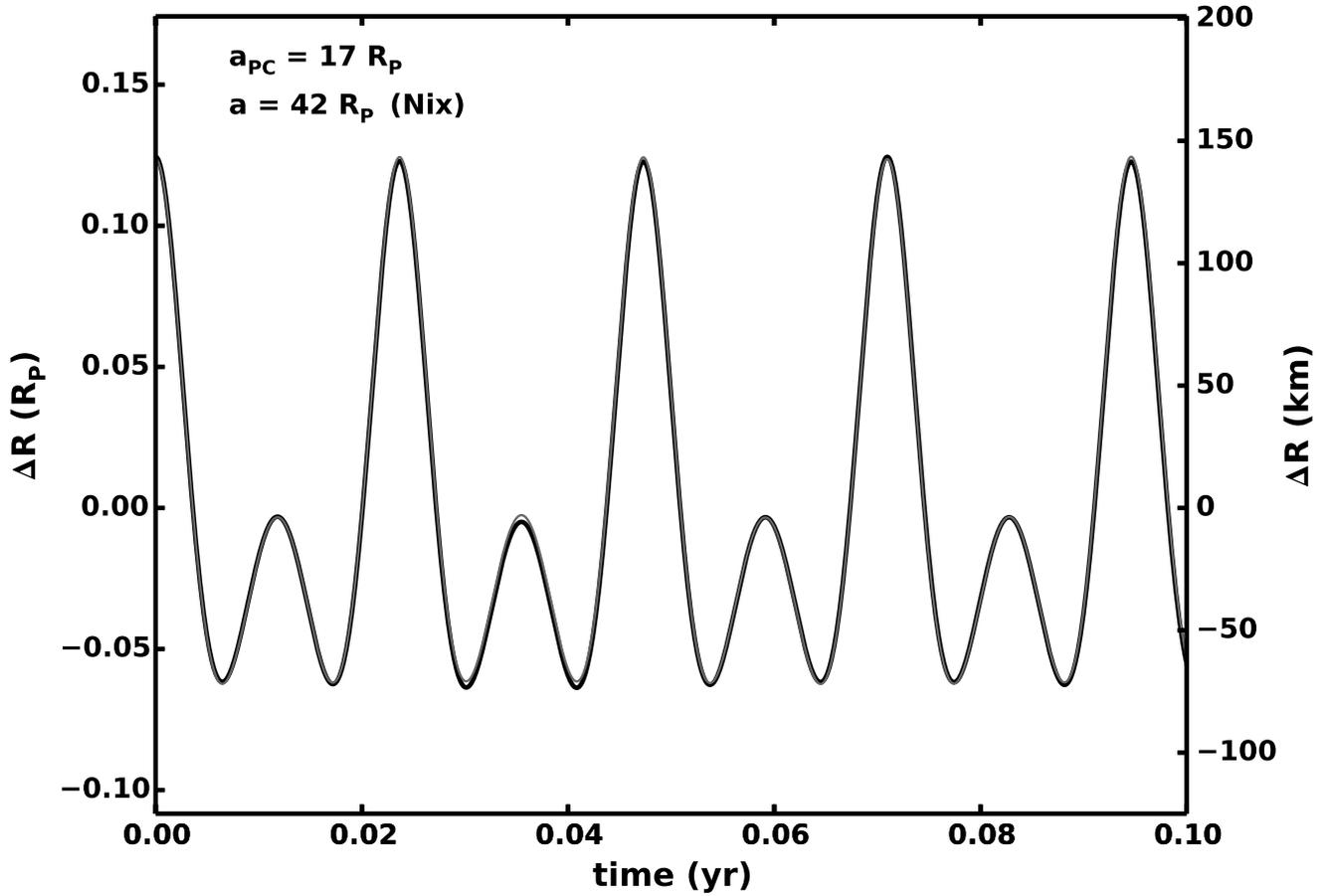}}
\caption{\label{fig:mostcircgen} The radial excursion of a ring
  particle on a most circular circumbinary orbit for a binary with
  $\ebin=0$. The darker curve is from points calculated with a
  simulation starting from a best-fit to a minimum excursion orbit,
  while the lighter curve is from analytical theory (the two curves
  follow one another very closely). The binary separation and orbital
  distance correspond to the satellite Nix around Pluto-Charon.  }
\end{figure}

Figure~\ref{fig:mostcircgen} provides an illustration of a particle on
a most circular orbit around a binary with $\ebin = 0$. It shows a
direct comparison between the analytical theory and a most circular
orbit found by numerically searching for the trajectory with minimum
radial excursions about its guiding center.  The small radial
excursions of a particle on a most circular orbit have a ``W'' shape
profile over a synodic period, showing (i) maxima when the satellite is
aligned with the binary and closest to the secondary, (ii) minima when
the satellite is perpendicular to the binary axis, and (iii) local maxima
when the satellite is closest to the primary.

In the circular binary's frame of reference, the orbit of a satellite
is a closed, egg-shaped loop.  All such figures are
nested. The further away from the binary, the more circular the figure.
Particles on most circular orbits traverse these paths on their own 
synodic period with respect to the binary. While their speeds may be 
different, their phases align in a way that all neighboring particles 
get coherently jostled by the binary. None of their paths cross.
\citet{pichardo2005} provide an excellent description of these
orbits in terms of `invariant loops', closed curves that vary in response
to the binary's potential \citep[see also][]{georgakarakos2015}.

\begin{figure}[htb]
\centerline{\includegraphics[width=7.0in]{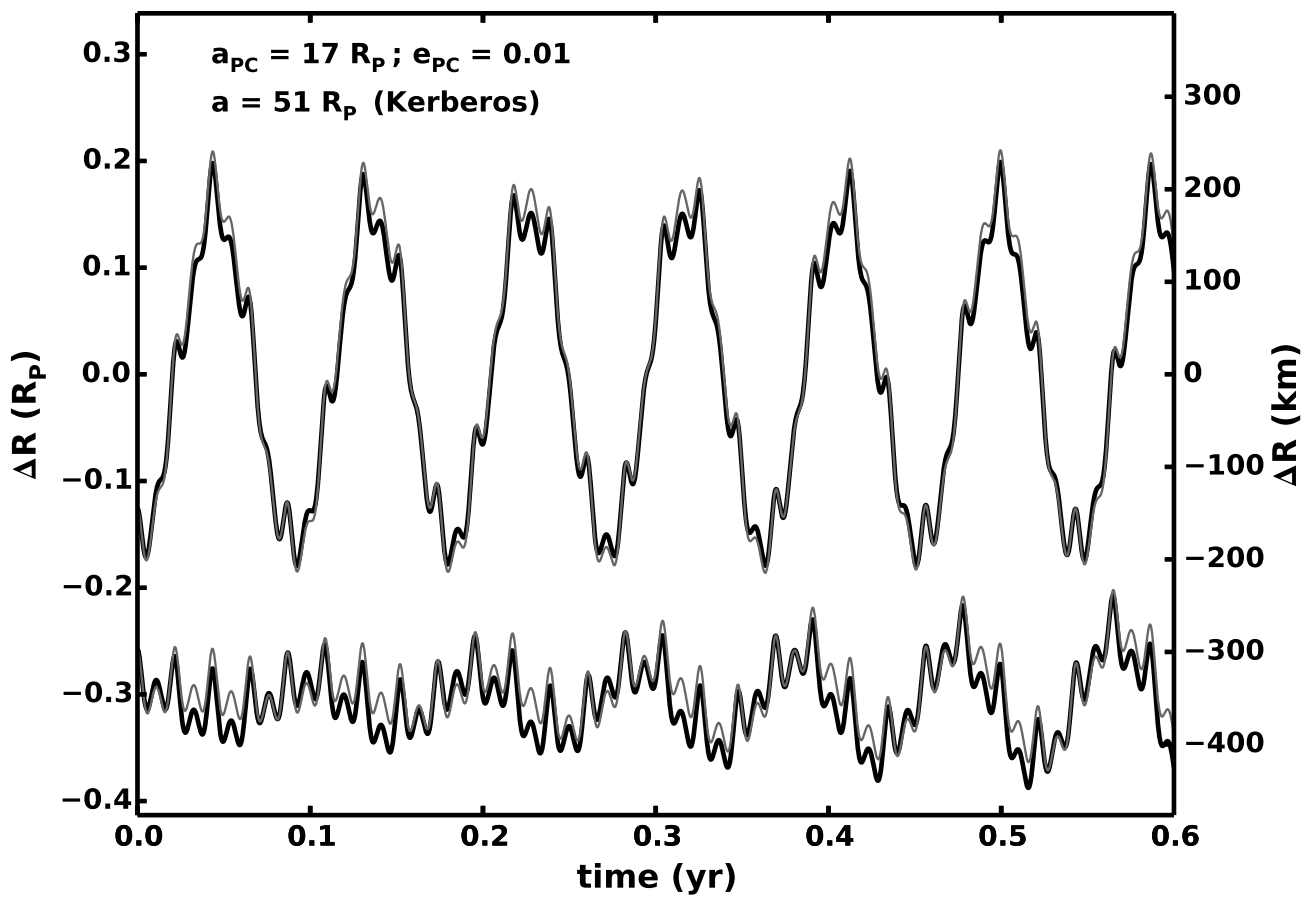}}
\caption{\label{fig:mostcircgenecc} The radial excursion of ring
  particles in orbit about a binary with $\ebin=0.01$. The upper dark
  curve is for a most circular path, while the lower dark curve
  (displaced vertically for clarity) is for a particle that has equal
  measures of free and forced eccentricity. In time, the growth of the
  lower curve will continue; it will produce a beat pattern with a
  maximum amplitude equal to twice that of the most circular curve
  \citep[e.g., Fig. 3 of][]{bk2015tatooine}.  The light shaded curves 
  are from analytical theory. The orbital elements are chosen to be 
  similar to Kerberos around Pluto-Charon.  }
\end{figure}

If the binary has some eccentricity, most circular orbits are
eccentric. Satellites following these paths have some epicyclic motion
about the guiding center, with eccentricity set by $\eforce$ in
Equation~(\ref{eq:eforce}). Figure~\ref{fig:mostcircgenecc} provides
an illustration, using a small binary eccentricity to allow the radial
excursion from the forced eccentricity to roughly match the radial
displacements oscillating at the synodic frequency.  With zero free
eccentricity, the radial excursions ($\sim\rgc\eforce$) are steady,
with no overall change in their maxima over time.

Figure \ref{fig:mostcircgenecc} also shows the effect of free
eccentricity.  Although the free eccentricity has nodal precession,
the forced eccentricity does not. Thus, the radial excursions relative
to a purely circular orbit about the binary center-of-mass vary
continuously, with a period $2\pi/|\ngc-\kappae|$.  Adopting a free
eccentricity equal to the forced eccentricity for the example in
Figure \ref{fig:mostcircgenecc}, the total eccentricity cycles between
zero and $2 \eforce$.  Thus orbits with no free eccentricity
($\efree=0$) have the smallest {\em maximum} radial excursion. More
importantly, these orbits do not intersect, offering a place in
phase-space for many small particles to coexist without colliding with
each other. 

\subsection{Summary}

We conclude this section with its main implication for ring dynamics.
The family of most circular orbits represents the dynamically cold
state of a ring orbiting a central binary. Collisional damping causes
particles to settle onto one of these orbits, just as for circular
orbits around a single central mass. As long as the ring is not
globally self-gravitating, the most circular orbits represent
reference frames in which the local evolution of the ring takes
place. Significantly, these orbit solutions exchange no net torque
with the binary.  Thus, the theory for ring dynamics developed for a
purely Keplerian potential applies to circumbinary orbits, as long as
the guiding centers are at distances where long-term stable orbits are
possible. 

Here we take advantage of these results by adopting reference frames
that track most circular paths. Then we may perform calculations such
as solving for the radial evolution of a circumbinary ring
(Equation~(\ref{eq:ringevolve})) simply by setting the orbital
distance variable $a$ to the guiding center radius. There are still
differences, although they are typically small. For example, the
viscosity in Equation~(\ref{eq:visc}) depends on the orbital period,
which is reduced by a factor of $ 3/8\times(\mu/M) (\abin/\rgc)$ in
units of the binary separation. At the orbit of Styx, this correction
is about 1\%.

The strategy of working in most circular frames fails at distances
close to binary or at the locations of resonant instabilities.
However, the effects of destabilizing multiple resonances
\citep[e.g.,][]{wisdom1980,lecar2001} are avoided if particles are
collisionally damped. Furthermore, as in Figure~\ref{fig:mostcircpcx},
damping and diffusion may facilitate resonant trapping and migration
\citep{ward2006}, processes that can enable a ring to expand as its
central binary evolves.  Preliminary estimates suggest that the
damping time expected for a sea of 1~km particles --- roughly 10~yr
--- is more than sufficient to mitigate the effects of the
instabilities outside of the 3:1 commensurability with the binary. On
their own, significantly larger objects, with radii exceeding 5~km or
so, may not be able to orbitally damp on their own, but instead
require a population of smaller bodies to keep them stable.

\section{The Pluto-Charon system}\label{sec:pc}

In this section, we apply the theory of circumbinary ring dynamics and
evolution to Pluto, Charon, and a ring where satellites form.  Aside
from deriving quantitative estimates for time scales and rates, we
numerically solve for the evolution of possible ring configurations.
We begin with a consideration of the binary itself, since its
evolution certainly affects the ring.

\subsection{Formation and evolution of the binary}

Pluto and Charon have radii $\RP \approx 1153$~km and $\RC = 603.6$~km
and masses $\MP \approx 1.305\times 10^{25}$~g and $\MC \approx
1.520\times 10^{24}$~g
\citep{buie2006,person2006,tholen2012,brozovic2015}.  Their orbit has
separation $\aPC \approx 16.97\ \RP$, eccentricity $\ePC$ consistent
with zero, and is tidally locked \citep{buie2012}.  The binary orbit
relative to the orbital plane of the major planets is significantly
inclined, with $\iPC \approx 96^\circ$.

An impact origin for the Pluto-Charon is plausible, both in terms of
the collision outcomes as compared to the binary's present configuration 
and the likelihood of such a collision in the outer solar system 
\citep{mckinnon1989,stern1992,canup2005,kb2014}. Simulations
\citep{canup2005} suggest that as much as $3\times 10^{21}$~g ---
about 0.02\% of the binary's mass --- ends up as debris around the
binary. This material initially lies on very eccentric orbits 
extending from distances of 5--10~$\RP$ to 30~$\RP$. If the orbits
circularize, the debris probably forms a compact ring within 15--30~$\RP$ 
of the center of mass \citep{kb2014}.  

\citet{kb2014} propose that collisional evolution of this ring leads
to the formation of the four known satellites and perhaps other,
smaller satellites at larger distances from Pluto-Charon. The known
satellites --- Styx (at an orbital distance of approximately $37~\RP$),
Nix ($43~\RP$), Kerberos ($50~\RP$) and Hydra ($57~\RP$) --- have a 
combined mass in the range of $3\times 10^{19}$~g to $3\times
10^{21}$~g \citep{buie2006,tholen2012,brozovic2015,showalter2015}. 
The combined mass is consistent with formation from debris left over 
from a collision which produces the Pluto-Charon binary. The orbital 
distances require expansion of the ring.

In an impact scenario, Pluto and Charon have an initial orbital separation of
\begin{equation} 
\aPC(t=0) \approx 5 \RP
\end{equation}
\citep{canup2005}.  The separation grew in time
through exchange of angular momentum that depends on the state of
matter within the two partners \citep[e.g.,][]{barr2015}; an idealized
theoretical rate describing the expansion is
\begin{equation}
\frac{d\aPC}{dt} = \kappa \aPC^{-11/2}
\end{equation}
where
\begin{equation} 
\kappa = 
3\frac{k}{Q} \frac{\MC}{\MP}\left[\frac{G(\MP+\MC)}{\RP^3}\right]^{1/2}
\RP^{13/2}
\end{equation}
with Love number $k = 0.058$ and a constant tidal dissipation function 
$Q = 100$ \citep[e.g.,][]{farinella1979,dobro1997,peale1999,cheng2014a}. 
The solution to these equations thus describes the orbital separation as
a function of time:
\begin{equation}\label{eq:pcevolve}
\aPC(t) = \min\left\{\left[13\kappa t/2 + \aPC(0)^{13/2}\right]^{2/13}, 
  17\RP\right\}.
\end{equation}
In this model, the binary expands by a factor of two within $10^5$~yr 
and reaches its present-day separation within 3~Myr, when Pluto and 
Charon become tidally locked.  

The initial eccentricity affects the tidal evolution,
typically by speeding up the expansion. However, for modest eccentricity,
the effect is weak, and in any event, the eccentricity damps relatively
quickly:
\begin{equation}
\frac{1}{\ePC}\frac{d\ePC(t)}{dt} \approx
-\frac{F}{3} \kappa \aPC^{-13/2}
\end{equation}
where the coefficient $F$ is $O(10)$ \citep{cheng2014a}.  The damping time
for an initial eccentricity of order unity in this case is
of the order of 1000~yr.  Hence, we assume that the binary eccentricity
$\ePC$ is zero unless otherwise specified.

\subsection{The Pluto-Charon ring: early conditions}\label{subsec:pcinit}

To describe the post-impact conditions in the Pluto-Charon system, we
define a fiducial model with a binary separation of $\abin = 5\ \RP$,
and circumbinary ring radius of $\aring = 20\ \RP$. We assume that the
mass of the ring is
\begin{equation}
\Mring = \Xring \cdot 3\times 10^{20} \ \textrm{g}
\end{equation}
where $\Xring \gae 1$ is a rough bound for our models in which the 
ring must contain enough mass to make the known satellites \citep{kb2014}.
We assume that this material has a Gaussian surface density profile 
with standard deviation of $\Delta a = 2\ \RP$:
\begin{equation}\label{eq:SigmaOh}
\Sigma(t=0) = \Sigma_G \exp\left[-(a-\aring)^2/2\Delta a^2\right] 
\end{equation}
\begin{equation}
\Sigma_G \approx 40 \Xring 
\aringXXRP^{\!-1}\!\DaIIRP^{\!-1}
\ \textrm{g/cm$^2$}.
\end{equation}
where $\Sigma_G$ is a constant relevant only to this Gaussian model.

To estimate plausible characteristic scales, we consider a
monodisperse population of ring particles with $\rp = 1$~km and $\rhop
= 1$~g/cm$^3$, and a characteristic surface density of 40~g/cm$^2$.
The number of particles in the ring is
\begin{eqnarray}
N & \approx & 7\times 10^4 \Xring \rpIkm^{-3}.
\end{eqnarray}
The speeds of individual particles that enter into ring dynamics are
\begin{eqnarray}
\label{eq:vkeppc}
\vkep = 0.21 \aXXRP^{1/2} \ \textrm{km/s} & \ \ \ \ \ \ \textrm{[orbital speed]}
\\ 
\label{eq:vescpc}
\vesc = 75\,\rpIkm \ \textrm{cm/s} & \ \ \ \ \ \ \textrm{[escape speed]}
\\
\label{eq:vhillpc} 
\vhill = 16\,\aXXRP^{\!-1/2}\rpIkm \ \textrm{cm/s} & \ \ \ \ \ \ \textrm{[Hill speed]}.
\end{eqnarray}
The latter two speeds are characteristic of gravitational stirring in the ring.

An important consideration for a compact ring is whether it is stable
against its own gravity \citep{toom64}. To avoid gravitational
clumping of particles within the ring, the particles must maintain a
minimum random velocity in addition to their orbital motion.  For
stability, the characteristic minimum speed is
\begin{equation}
  \vstab = 0.94 ~ \aXXRP^{3/2} \SigmaXLcgs\ \textrm{cm/s}.
     \ \ \ \ \ \textrm{[minimum stability speed].}
\end{equation}
In a ring of identical icy particles, this condition translates to a
radius of about 10 meters if they stir each other to their escape
speed. Since the ring expands to larger radii and lower surface
densities, the minimum speed and radius both fall as the ring
evolves. Furthermore, if the impact debris contains larger embedded
fragments, it will settle to higher speeds, comparable to the escape
velocity of those fragments. Thus the bulk of the ring material can be
composed of smaller particles and yet the ring will remain gravitationally
stable.

In addition to the binary separation and the orbital radius of the ring,
there are several other useful length scales:
\begin{eqnarray}
\rroche \approx 3700 ~ \textrm{km} \approx 3 \Rp & 
      \ \ \ \ \ \ \ \ \ \textrm{[Roche limit]}\textrm{\Huge \ }
\end{eqnarray}
\begin{eqnarray}
\rhill = 11 \aXXRP\rpIkm \ \textrm{km} & \ \ \ \ \ \textrm{[Hill radius]}
\\
h = 84 \aXXRP^{3/2}\vpLXXVcms \ \textrm{km} & 
\ \ \ \ \ \textrm{[ring scale height]}
\end{eqnarray}

Finally, results from \S2 provide relevant time scales. Dynamical times
are
\begin{eqnarray}
\tbin & = & 1.0 \aPCVRP^{3/2}\ \textrm{d} \ \ \textrm{[binary orbital period]}
\\ \tkep & = & 8.2 \aXXRP^{3/2}\ \textrm{d} \ \ \textrm{[ring orbital period]}
\end{eqnarray}
Ring evolution time scales are
\begin{eqnarray}
\label{eq:tdamppc}
\tdamp & = &
14\,
\aXXRP^{3/2}\!\rpIkm\!\flossOpVI\!\SigmaXLcgs^{-1}
\ \textrm{yr}
\\
\tmig & \gae & 180,000\ 
\!\aXXRP^{\!-1/2}
\!\rpIkm^{\!-1}
\!\SigmaXLcgs^{\!-1}
 \ \textrm{yr}
\\
\label{eq:tspreadpc}
\tspread & \sim & 450,000\ 
\!\aXXRP^{\!1/2}
\!\rpIkm^{\!-1}
\!\SigmaXLcgs^{\!-1}
\!\vvesc^{\!-2}
 \ \textrm{yr},
\end{eqnarray}
where the migration time $\tmig \sim a/\dadtweakgrav$, and $\tspread$ is
$(a/2)^2/\nu$, giving a rough time scale for the ring to double its
size by viscous diffusion.  The gravitational spreading time scale is
a crude estimate in the limit of a dynamically cold disk; dynamically
hot rings, with scale heights that greatly exceed the particle Hill
radii, will experience slower evolution by gravitational scattering.
The numerical value for the viscous diffusion time carries the
assumption that the random component of the particle velocities
(governing the ring viscosity) are given by the escape velocity from
the particle surface. The surface density $\Sigma$ refers to the local 
surface density in the ring.

For a ring of monodisperse particles with random velocities comparable
to particle escape velocities, damping is generally much faster than 
spreading.  For small particles ($\rp < 1$~km) in low-mass rings 
($\Sigma < 40$~g/cm$^2$), evolution time scales are comparable to
the expansion time of the binary ($10^5$--$10^6$~yr).  Large particles 
or large ring mass can shorten the time for the ring to spread. However,
it is hard to reduce the spreading time significantly unless particle
sizes reach 50--100~km. The small number of ring particles then makes
spreading unlikely \citep[see also][]{peale2011,cheng2014b}.

If the ring consists of particles with a broad range of sizes, viscous 
diffusion can drive a rapid expansion of the ring. If a few large 
particles (radius $r_{\rm large}$) are embedded in a sea of small ones 
(radius $r_{\rm small}$), gravitational stirring by the large particles
drives the random velocities of the smaller particles to the escape 
velocity of the large particles. If the large particles have sufficiently 
large escape velocities, the formal spreading time is very short
\citep{gold2004}:
\begin{equation}
\label{eq:tspreadstirredpc}
\tspread  \sim  5\ 
\!\aXXRP^{\!1/2}
\!\rpsmallIm
\!\rplargeXkm^{\!-2}
\!\SigmaXLcgs^{\!-1}
 \ \textrm{yr},
\end{equation}

This expression assumes that the rate of collisional damping for the
small particles is small compared to the stirring rate of the large
particles. For a swarm of 1~m particles, the likely damping time of
0.01~yr (Equation~(\ref{eq:tdamppc})) is smaller than this spreading
time. However, rapid damping promotes the growth of small objects into
larger ones \citep[e.g.,][]{kb2014}, slowing the rate of damping and
accelerating the rate of spreading.  Because the damping time grows
much faster with $a$ than the spreading time, it seems plausible that
a swarm of particles with sizes ranging from 1~m to 1--10~km could
spread radially on time scales ranging from a century to many
millenia.

While a two-population model for the ring is simplistic, it captures
the relevant physics and has been used successfully in theories for
planet formation in circumstellar disks
\citep[e.g.,][]{weth1989,ida1993,gold2004}.  For Pluto-Charon, the
early evolution of the debris is likely dominated by destructive
collisions of particles on very eccentric orbits. For typical
collision velocities of 0.1--0.2~km/s (Equation~(\ref{eq:vkeppc})),
center-of-mass collision energies of $10^7 - 10^8$ erg/g destroy
particles with $r \lesssim$ 5--10~km
\citep{benz1999,lein2012,kb2014}. Large self-gravity saves larger
particles from significant mass loss.  Rapid collisional damping likely
saves much smaller particles, $r \lesssim$ 1--10~m, from destruction
\citep{kb2014}.  Fragmentation likely converts intermediate-mass
particles, 1--10~m $ \lesssim r \lesssim$ 5--10~km, into smaller
particles which damp rapidly. Thus, two populations of particles ---
small particles with most of the mass and a few large particles with
significant self-gravity --- are plausible components of a circularized
ring of particles around Pluto-Charon.

\subsection{Evolution of the Pluto-Charon ring}

To investigate the evolution of monodisperse and two-population
ensembles of particles in a ring orbiting Pluto-Charon, we solve the
radial diffusion equation (Equation~(\ref{eq:ringevolve})) using a
finite difference code \citep[see also][]{bk2011a}.  The code tracks
the surface density $\Sigma$ in $O(10^3)$ radial bins spanning $a =
5\,\RP$ out to $120\,\RP$. In each bin, we calculate the viscosity
(Equation~(\ref{eq:visc})) and the radial mass flux from gravitational
scattering (Equation~(\ref{eq:dadtweakgrav})). For each time step, the
finite difference code updates the surface density from the viscosity
and radial mass flux.  Tests on standard problems
\citep[e.g.,][]{lbp1974,bk2011a} verify the accuracy of our solutions
to better than 1\% over $10^6 - 10^7$ yr.

In applying our code to the circumbinary environment, we adopt the
results of \S\ref{sec:circumbin}.  Damping and stirring of particles,
along with radial transport, take place relative to most circular
orbits. Hence, our use of orbital distance $a$ is synonymous with most
circular orbit positions characterized by the orbital radius of the
guiding center $R_g$.  We also assume that resonant excitations and
instabilities within the ring do not contribute significantly to the
evolution.  In our fiducial ring with 1~km particles, the damping time
is rapid enough to suppress resonant-driven excitations.  These
idealizations still allow us to reach our goal of understanding how the
ring might spread as the central binary expands.

To establish the range of possibilities for the expansion of a
compact ring around a young Pluto-Charon, we choose from a palette of
models to test the effects of particle size ($\rp$),
velocity ($\vp$), and total mass in the ring ($\Mring$). Our initial 
conditions assume a Gaussian surface density
profile (Equation~(\ref{eq:SigmaOh})) with $a = 20\,\RP$, $\Delta a =
2\,\RP$. We start the binary at an orbital separation of 5~$\RP$,
and have it expand on a time-scale of $\sim 10^5$~yr, according to
Equation~(\ref{eq:pcevolve}).

An additional model parameter is the choice of the inner boundary
condition. We can set the inner boundary to be ``flow-through'' under
the assumption that all material reaching the inner edge is lost,
presumably to ejection or accretion by the binary
\citep[e.g.,][]{pri1991}.  However, in most of our trials, the inner
boundary is reflective at the edge of the stability zone.  We enforce
this boundary by setting up an artificial outflow that is constant
inside the ring edge and falls off exponentially over a scale length
of 2.5\% of the orbital distance of the edge.  This algorithm prevents
a strong discontinuity from building up at the inner edge of the
ring. Even with this approach, the gradient at the edge can be large
and can lead to numerical instability. Thus, we add artificial
viscosity in the vicinity of the edge in cases where gravitational
scattering strongly dominates over viscous spreading.

For all models we set $\Sigma$ to zero at the outer boundary, although
in most cases the ring never reaches the outer edge of the
computational domain.

The following list summarizes our models. We use the approximate
time for the ring to reach 60~$\RP$, just beyond 
the present-day orbit of Hydra, as a benchmark. If this time 
is within $\sim 10^4$ yr, then the ring may spread before
large satellites grow within it \citep{kb2014}. Otherwise,
some evolution of particle size --- or some other
means of migration --- is required as the ring expands.

\begin{figure}[htb]
\centerline{\includegraphics[width=7.0in]{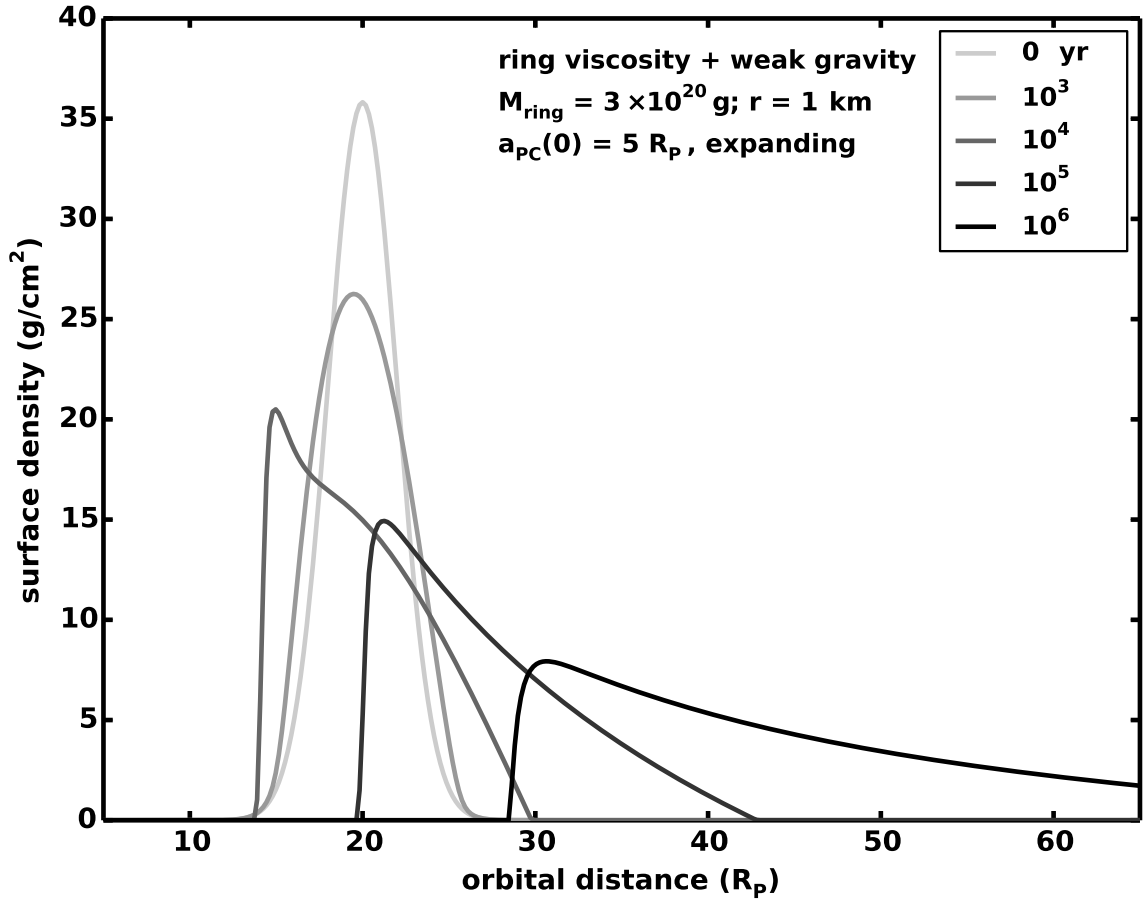}}
\caption{\label{fig:spreadPC} Spreading of a circumbinary ring. The
  curves show snapshots of the surface density as a function of
  orbital distance from Pluto-Charon. The rings spread from viscosity
  and gravitational scattering as the binary expands. The binary
  pushes the inner edge of ring outward, keeping it at a distance of
  twice the binary separation.  }
\end{figure}

\begin{figure}[htb]
\centerline{\includegraphics[width=7.0in]{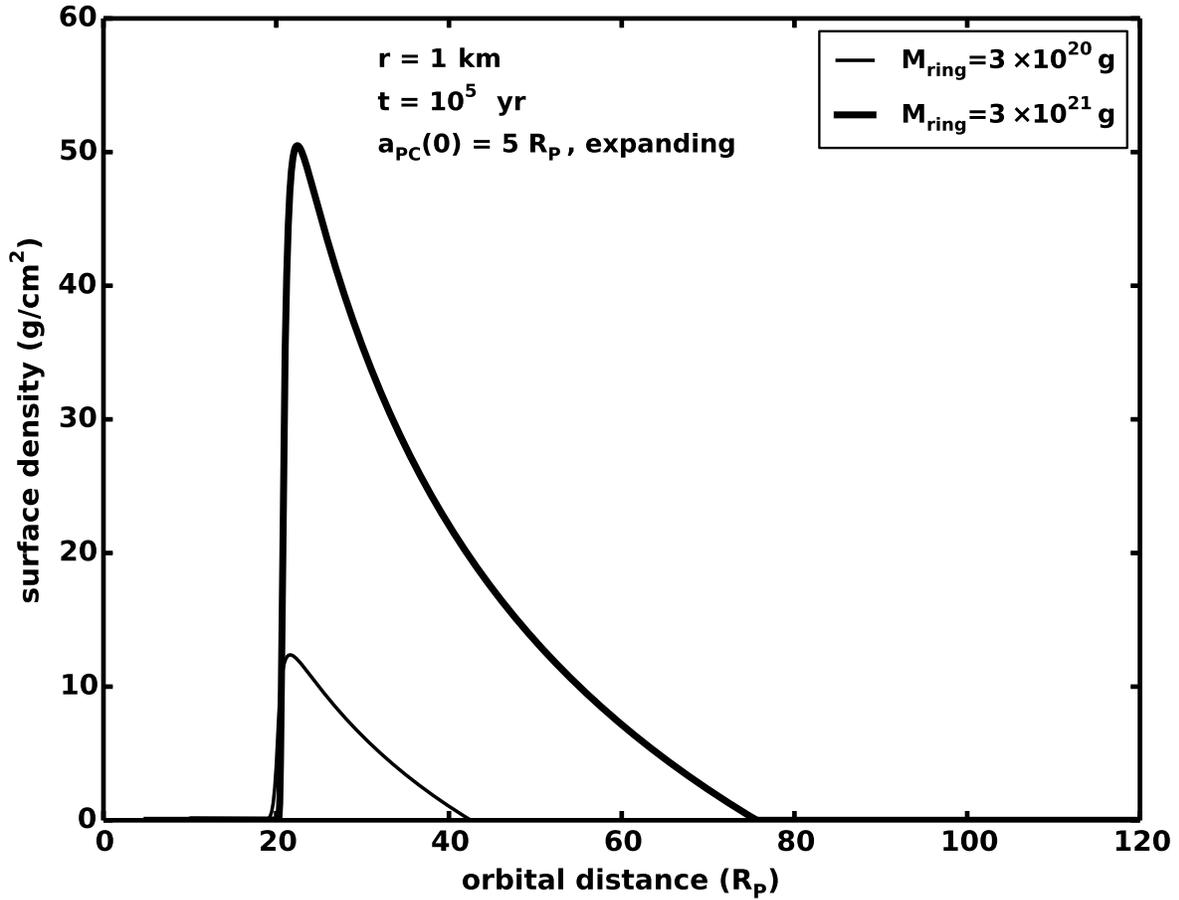}}
\caption{\label{fig:spreadPCmring} 
Spreading of circumbinary rings of two different masses
as in the previous figure but with profiles shown only at $10^5$~yr.
The rings initially differ only in their surface density; the outcomes
show how the increased surface density drives the spreading.
}
\end{figure}

\begin{figure}[htb]
\centerline{\includegraphics[width=7.0in]{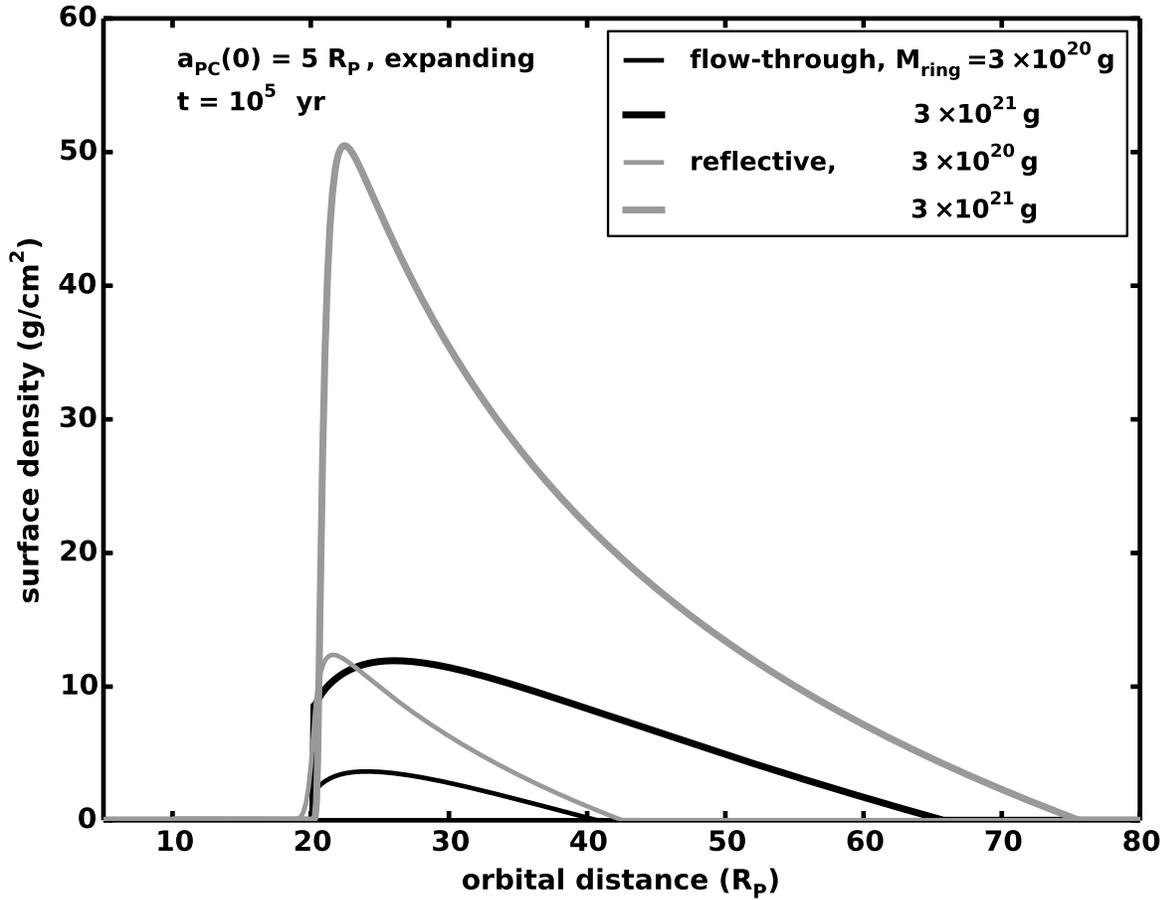}}
\caption{\label{fig:spreadPCleak} 
Spreading of a massive circumbinary ring with mass loss at 
the inner edge. The curves show the effect of mass loss
at the inner edge of the ring, which expands along with the
binary. 
}
\end{figure}

\begin{figure}[htb]
\centerline{\includegraphics[width=7.0in]{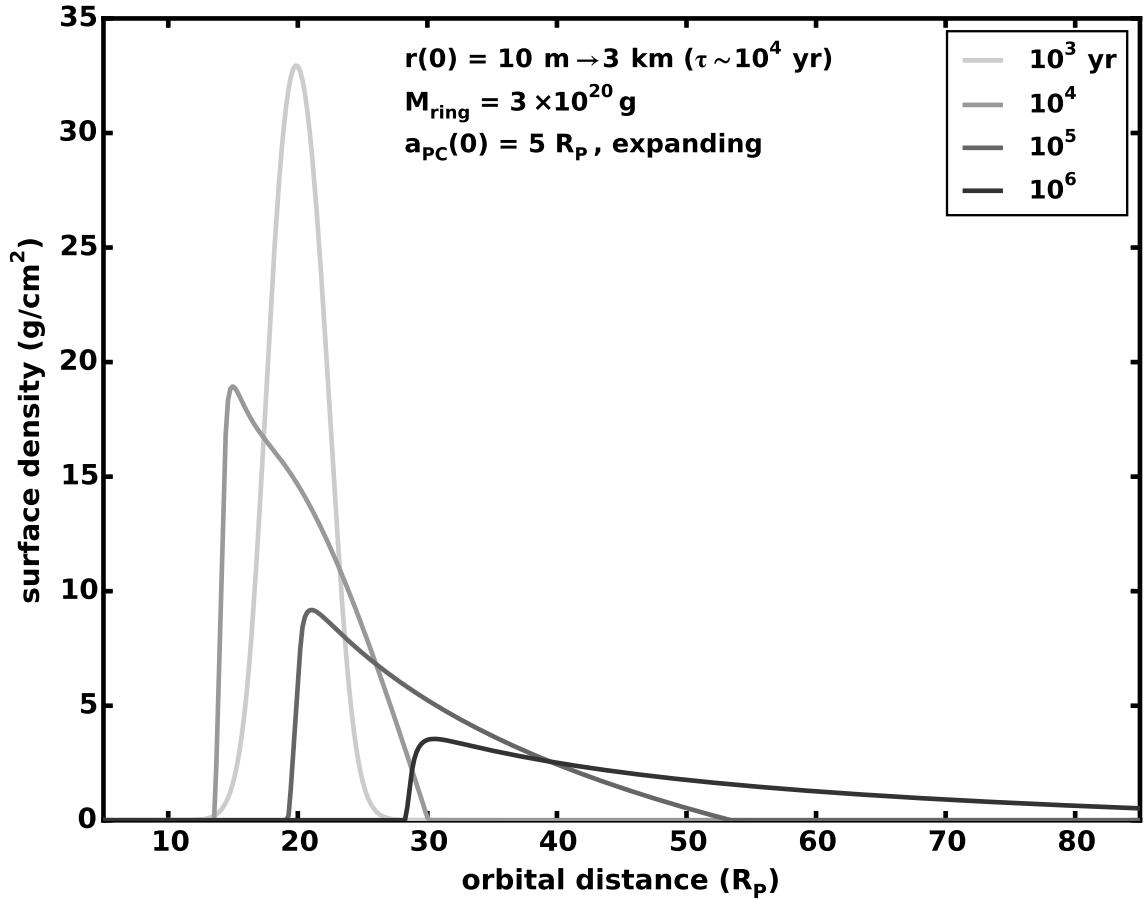}}
\caption{\label{fig:spreadPCgrow} The spreading of a circumbinary ring
  with growing particles as in Figure~\ref{fig:spreadPC} except
  the mass grows (along with the particle speeds, set by the
  escape velocity) from 10~m to 3~km on a time scale of $10^4$~yr. 
  The growth allows for modest spreading beyond the baseline
  model with particle radii fixed at 1~km.}
\end{figure}

\begin{figure}[htb]
\centerline{\includegraphics[width=7.0in]{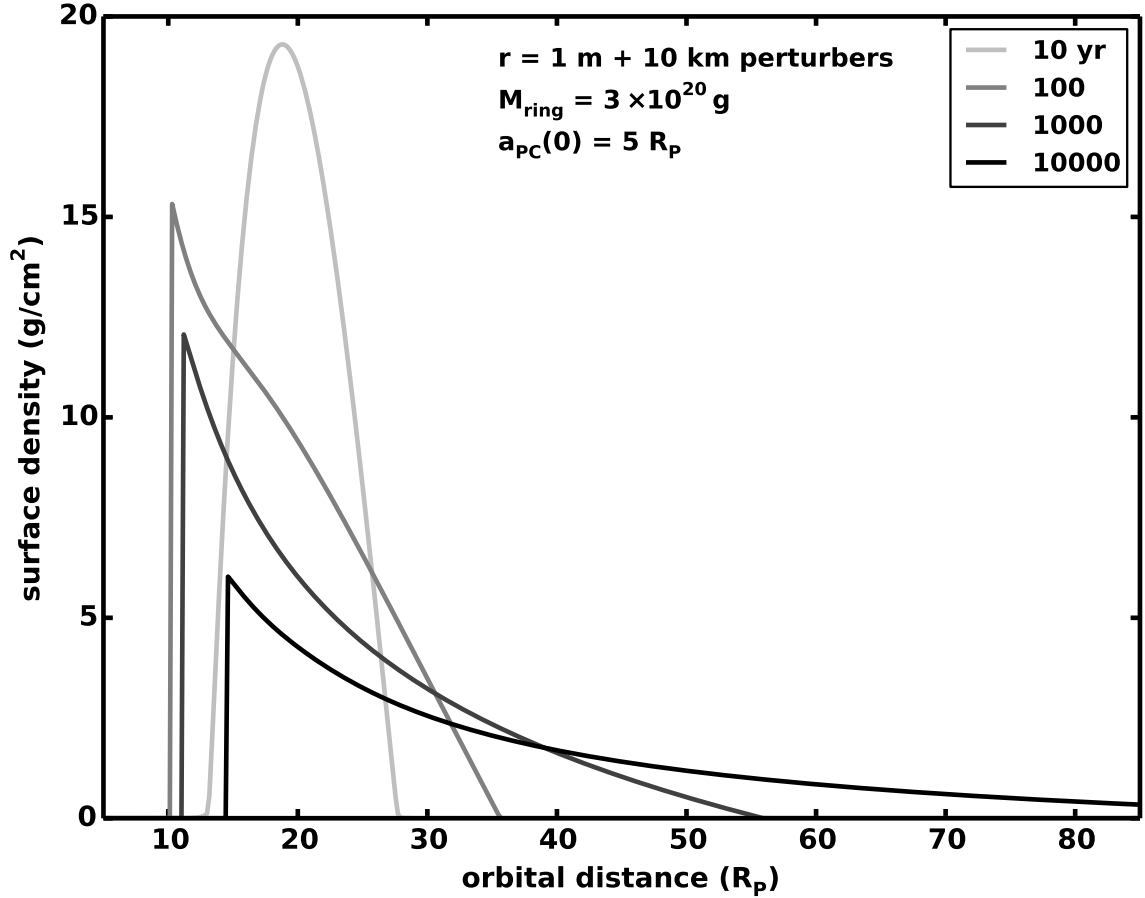}}
\caption{\label{fig:spreadPCrp1mleak} Rapid spreading of a
  circumbinary ring of small particles with embedded satellites. A
  population of meter-size objects stirred to 0.5 times the Hill
  velocity of 10-km objects (evaluated at $20\ \RP$) spreads rapidly,
  as seen in the sequence of curves. In this case, material can flow
  through the inner edge of the ring. While this model is an
  idealization, it shows how effective viscous spreading can be when
  the relative velocities are pumped up.  }
\end{figure}

\begin{itemize}

\item{\em Baseline model.} Our starting point is a ring with total
  mass $3\times 10^{20}$~g ($\Xring = 1$) made up of kilometer-size
  particles stirred to their escape speed ($\sim$75~cm/s).
  Figure~\ref{fig:spreadPC} illustrates the spreading of this ring.
  This ring spreads slowly compared to satellite growth times,
  extending past $60\ \RP$ in a few times $10^5$~yr.

\item{\em Massive rings.}  If the ring is more massive than the 
baseline model, the spreading time is reduced 
  ($\tspread \sim 1/\Sigma \sim
  1/\Xring$). Figure~\ref{fig:spreadPCmring} gives an example for a
  ring with $\Mring = 3\times 10^{21}$~g ($\Xring = 10$). In this model,
  the ring reaches 40~$\RP$ in a $10^4$~yr and is beyond 70~$\RP$ by
  $10^5$~yr. Compared to the estimated growth time scale of $10^3-10^4$~yr,
   this expansion rate is still slow.

\item{\em Massive rings with loss through the inner boundary.}  In
  previous models, we use reflective inner boundaries, since material
  at the inner edge is presumably pushed back into the ring by the
  binary. Here we explore the possibility that mass instead flows
  through the inner edge and is ejected or accreted by the
  binary. Figure \ref{fig:spreadPCleak} shows an example. It shows the
  baseline model, as well as a model with a ring that is ten times
  more massive, both with and without loss of mass through the inner
  boundary. The effect of the mass leakage is a loss of about 20\% of
  the total mass, a flattening of the surface density profile, and a
  slightly less extended disk.

\item{\em Growth of particles.} We start with the baseline model,
  but with particles of radius $\rpmin = 10$~m.  These particles
  grow monotonically to a maximum radius of $\rpmax = 3$~km on a time
  scale of $\tau_{\rm grow}$ according to
\begin{equation}
\rp(t) = \rpmin + (\rpmax-\rpmin) \left[1 - \exp(-t/\tau_{\rm grow})\right].
\end{equation}
The random velocity is set to be the instantaneous escape speed of the
particles. Figure~\ref{fig:spreadPCgrow} shows the result. It differs
from the baseline model because the spreading starts off more slowly
and becomes more rapid only at the late stages, as $\rp$ grows. Even so,
the ring extends to only 50~$\Rp$ in $10^5$~yr.

\item{\em Stirring by embedded satellites.}  Guided by the physics of
  planet formation \citep{kb2004a}, we finally consider a scenario in
  which a population of small ($\rp=1$~m) particles is stirred by a
  few larger ones ($\rp = 10$~km).  
  We include the effect of these larger bodies by stirring the smaller
  ones to a random speed of half of the Hill velocity of 10~km objects
  (fixed at its value at $a=20\,\RP$). This choice acknowledges that
  the larger bodies are rare, and that damping among the smaller
  bodies prevents them from being stirred to the escape speed of
  the embedded satellites. The result is presented in
  Figure~\ref{fig:spreadPCrp1mleak}. The ring expands to Hydra's
  present-day orbit in just over 1000~yr, and by $10^4$ yr, it
  has spread to well over 100~$\RP$.
\end{itemize}

To summarize our results, monodisperse populations of ring particles
stirred to their own escape velocity, have a spreading time that
decreases with particle radius and mass of the ring (Equation
(\ref{eq:tspreadpc})). For particles with radii larger than a few
kilometers and for ring masses larger than $\sim 10^{21}$~g, the ring
spreads to Hydra's orbit within $10^5$~yr.  Allowing for particle size
to grow from some small value ($\rp \sim 10$~m) only delays the
spreading.

In models inspired by planet-forming disks, with small particles and
embedded satellites, the ring can spread quickly (Equation
(\ref{eq:tspreadstirredpc})). Even a low mass ring ($3\times
10^{20}$~g) can spread out past Hydra on a time scale of $\sim
1000$~yr. A weakness of this scenario is that the stirring is kept
at a constant level throughout the ring as it spreads.  For this
scenario to be possible, embedded satellites must either migrate
within the disk, or form in the outer regions of the disk as it
spreads. This latter possibility seems plausible, since formation
times can be short if the ring should dynamically cool by collisional
damping \citep{kb2014}. 

\section{Conclusions}

We consider the evolution of a ring of particles around a central
binary, with a focus on the Pluto-Charon system. Compared to earlier
theories for evolving disks of particles
\citep[e.g.,][]{cook1964,gold1978}, we add two new features: (i)
gravitational scattering, which we quantify in terms of weak
scattering theory, and (ii) the impact of the circumbinary environment
on ring evolution.  Gravitational scattering by massive particles can
contribute significantly to the expansion of the ring. As for the
circumbinary environment, aside from defining an orbital plane and
creating an inner edge by destabilizing orbits \citep{holman1999}, the
impact of the binary is surprisingly small.

Our main results are listed here, followed by a summary of our conclusions.

\begin{figure}[htb]
\centerline{\includegraphics[width=7.0in]{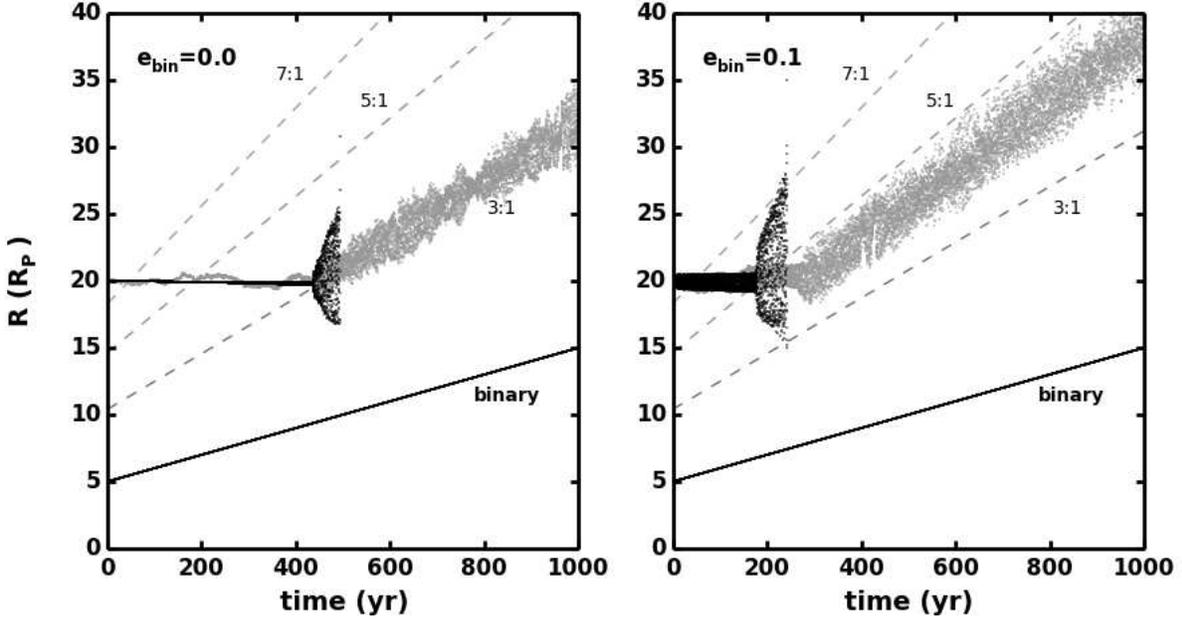}}
\caption{\label{fig:mostcircpcx} The orbital evolution of a satellite
  around an expanding binary.  The left panel shows a circular
  ``Pluto-Charon'' binary expanding from an orbital separation of
  5~$\RP$ (lower black curve). The dashed lines indicate the location
  of the 3:1, 5:1, and 7:1 commensurabilities expanding with the
  binary. A satellite on its own (black dots on the far left) orbits
  at a radial distance of 20~$\RP$ until it is overtaken by an
  instability near the 3:1 commensurability at about 400 yr. Then
  it is ejected from the system over a period of hundreds of orbits. A
  satellite that experiences collisional damping and radial diffusion
  (gray dots) survives the instability and ``surfs'' the 
  3:1 commensurability as it expands. The damping and diffusion
  parameters are drawn from fiducial values in \S\ref{subsec:pcinit}
  ($\rp = 1$~km, $\vp = \vesc$, $\Sigma = 40$~g/cm$^2$). The
  right panel is similar, but with an eccentric binary ($\ebin =
  0.1$). In this case, an isolated satellite (black dots) is unable 
  to navigate the 5:1 commensurability.  A damped, diffusing satellite
  survives and rides resonances near the 4:1 commensurability.}
\end{figure}

\begin{itemize}

\item\textbf{Ring evolution.}  Our analysis of weak gravitational
  scattering provides a natural extension to the viscous evolution of
  particle rings and disks. The expressions of \citet{cook1964} and
  \citet{gold1978} hold until gravitational interactions start to
  become important.  The demarcation between the viscous and gravity
  regimes is the ratio of the random speeds of particles to their
  typical escape speed. If the random speeds are large relative to
  $\vesc$, the particles do not feel their mutual gravity --- evolution
  is viscous. When particles orbit beyond the Roche limit and have
  random speeds comparable to or smaller than their escape velocities,
  they stir one another gravitationally.  Our results allow a smooth
  transition between these two regimes.

\item\textbf{Most circular orbits.}  The circumbinary environment
  regulates the inner edge of a ring and can destabilize orbits near
  close-in resonances
  \citep[e.g.,][]{wisdom1980,murray1999,pichardo2008,popova2013}.
  Although the binary's time varying gravitational potential modifies
  orbits in the rest of the ring, the family of most circular orbits
  provides reference frames similar to the circular orbits around a
  single central mass \citep{lee2006, leung2013}. Dissipative disks or
  rings settle on to these orbits, and perturbations (e.g., from
  gravitational stirring) cause particles to acquire free eccentricity
  and inclination relative to these frames.  Thus, we can calculate
  ring evolution in the same way around a binary as we do around a
  single planet or star \citep{bk2015tatooine}.

\item\textbf{Secular excitations.} In initiating the research reported
  here, our expectation was that stirring by the binary might help the
  viscous spreading of the ring
  \citep[e.g.,][]{mori2004,paarde2012,raf2013,lines2014}.  However,
  collisional damping causes particles to settle on most circular
  paths, even in the presence of a massive axisymmetric disk and
  distant massive perturbers. Thus, the central binary is not
  necessarily a source of stirring or torque exchange. We explore the
  implications for the early stages of circumbinary planet formation
  in a separate paper \citep{bk2015tatooine}.

\item\textbf{Resonance effects and instabilities.}  The binary can
  still make a mess of orbits.  In units of the binary separation, the
  Pluto-Charon satellites are closer in than known circumbinary
  planets from NASA's \textit{Kepler} mission
  \citep[e.g.][]{doyle2011,welsh2012,orosz2012a,orosz2012b}. And,
  unlike most of the exoplanets, the Pluto-Charon moons orbit inside of
  suspected resonance-driven instabilities. If the moons emerge early
  in a compact ring, then they can get caught up in resonant orbits
  that grow along with the tidally evolving binary
  \citep{ward2006}. While they may be transported in the resonances,
  they may just as easily get ejected, as our own simple experiments
  demonstrate (Figure~\ref{fig:mostcircpcx}).  For the Pluto-Charon
  system, coordinating resonance transport in a way that preserves all
  of the moons does not seems possible \citep{cheng2014a}.

\item\textbf{Collisional damping can stabilize resonant orbits.}  Our
  simple simulations also demonstrate how collisional damping and
  diffusion can protect satellite orbits from ejection \citep[see
    also][]{ward2006,lith2008b,cheng2014a}.  Thus a ring of small
  particles can survive the sweep of resonances during tidal
  expansion. Similarly, as long as a substantial reservoir of small
  particles is in place, we speculate that between collisional damping
  among the small objects and dynamical friction between these small
  bodies and the large ones, the entire population is protected agains
  unstable resonances.

\item\textbf{Evidence for resonant migration.}  Our simple simulations
  also demonstrate how collisional damping and diffusion can
  facilitate resonant migration \citep{ward2006}. Small particles
  become trapped near resonances and then get swept up in them as 
  the binary expands. Figure~\ref{fig:mostcircpcx} illustrates the
  phenomenon and highlights the role of binary eccentricity in the
  process.  If a reservoir of small particles is maintained during
  the binary's tidal evolution, small moons might also participate 
  in this migration.  It is beyond our scope here to explore these 
  possibilities in detail, but our anecdotal results are promising.
  We plan more detailed simulations to test this possibility.

\end{itemize}


On the basis of these ideas and our numerical simulations,
we infer the following conclusions for the Pluto-Charon system:

\begin{itemize}
\item Starting with a compact ring around Pluto-Charon with mass
  comparable to the observed moons, viscosity and gravity can
  spread the ring to the moons' present locations in less than 
  $10^5$--$10^6$~yr. These time scales are comparable to the tidal
  evolution of the binary's semimajor axis, but are substantially
  longer than the growth time of 10--100~km satellites ($\sim
  10^3$~yr).

\item The spreading of the ring can be comparable with, or faster
  than, the growth time if (i) the ring has small (1-meter) particles
  stirred by large (1--10~km) bodies, or (ii) the mass of the ring is
  large ($\gae 10^{22}$~g). In scenario (i) large objects can probably 
  form as the ring spreads.  In both scenarios, the ring spreads
  quickly to distances well beyond Hydra, the outermost known moon.

\item Resonance-driven instabilities remove isolated satellites or
  moons \citep[e.g.,][]{cheng2014a}. This phenomenon is complicated by
  the tidal evolution of Pluto-Charon, which causes commensurabilities
  to sweep through the system. However, collisional damping and
  diffusion can make satellites immune to resonance-driven
  instabilities and can also facilitate trapping in a resonance. In
  the latter case, ring material can be transported outward along with
  the resonances, as envisioned for full-fledged moons by
  \citet{ward2006}.

\item Together, the spreading mechanisms and interactions with
  resonances suggest potential growth modes for the
  satellites. \textit{In situ} formation can occur quickly
  \citep[10$^3$ yr;][]{kb2014}; the time needed for spreading the ring
  can be comparably fast or slow (10$^5$~yr), leading to these
  possibilities:
  
  \begin{enumerate}[i.]
  \item The spreading might occur before the expansion of the binary.
    Then the sweep of resonances might eject or collisionally destroy
    the largest satellites. They would have to re-emerge out of the
    debris once the sweep was complete. This scenario would require a
    massive initial ring, as particles could be lost in the process.
    (\citealt{kcb2014} describe a similar picture for Fomalhaut b.)
    Simulations suggest that there was enough mass in the ring
    initially to sustain a mass loss of over 99\% and still have
    enough to make the moons \citep{canup2005,canup2011}. The result
    could be an extended disk and small as-yet-undetected moons beyond
    Hydra \citep{kb2014}.

  \item If the ring spreads slowly, then particles in it could get
    caught up in one or more resonances with the expanding
    binary. \textit{In situ} formation during the expansion is
    possible, but a reservoir of small particles (perhaps replenished
    by collisions) would be essential to prevent the instabilities
    from leading to ejection. Preliminary numerical experiments
    (Figure~\ref{fig:mostcircpcx}) show that material might be trapped
    in several different commensurabilities, depending on the binary
    eccentricity and the particle damping/diffusion.
  \end{enumerate}
\end{itemize}

The next step is to incorporate the physics of ring spreading into
{\it Orchestra}, our hybrid N-body--coagulation code
\citep{bk2011a}. Then, we can track the detailed interplay between
ring spreading and satellite growth.  The N-body part will be
essential to simulate resonant-driven instabilities that can clear out
orbits in the ring \citep{cheng2014b}.  How these instabilities play
out will depend on a number of factors including the effectiveness of
orbital damping, and the time-varying eccentricity of the binary,
since circularization brings some stability \citep{holman1999}. Our
code will enable us to determine how collisional damping,
gravitational stirring, resonance-driven instabilities, ring spreading
and tidal evolution can work in concert to make the binary's intricate
system of moons.

Meanwhile, our models suggest that the expansion of a narrow ring
around the Pluto-Charon binary is plausible. The extent of the expansion
depends on the details. The presence of small moons beyond Hydra's
orbits would suggest a more massive disk with small particles stirred
by larger ones. Then, we expect 3--5 satellites with radii of a
kilometer or so, on circular orbits beyond 60~$\RP$ \citep{kb2014}.
Ring evolution models with less mass and less stirring leave the ring
truncated at Hydra's orbit.  Even so, scattering by the growing moons
could place some satellites beyond Hydra, although their eccentricity
is expected to be significant.  \newhorizons, with its unprecedented
view of the Pluto-Charon satellite system, will test these scenarios.

\acknowledgements

We are grateful for the comments of an anonymous referee that resulted
in a number of improvements to our presentation.  We thank NASA for support
through the \textit{Astrophysics Theory} and \textit{Origins of Solar
  Systems} programs (grant NNX10AF35G) and through the \textit{Outer
  Planets Program} (grant NNX11AM37G). We are also grateful to NASA
for a generous allotment of supercomputer time on the 'discover'
cluster.

\bibliography{planets}{}

\begin{thebibliography}{}
\expandafter\ifx\csname natexlab\endcsname\relax\def\natexlab#1{#1}\fi

\bibitem[{{Barge} \& {Pellat}(1990)}]{barge1990}
{Barge}, P., \& {Pellat}, R. 1990, \icarus, 85, 481

\bibitem[{{Barr} \& {Collins}(2015)}]{barr2015}
{Barr}, A.~C., \& {Collins}, G.~C. 2015, \icarus, 246, 146

\bibitem[{{Benz} \& {Asphaug}(1999)}]{benz1999}
{Benz}, W., \& {Asphaug}, E. 1999, Icarus, 142, 5

\bibitem[{{Brahic}(1976)}]{brah1976}
{Brahic}, A. 1976, Journal of Computational Physics, 22, 171

\bibitem[{{Bridges} {et~al.}(1984){Bridges}, {Hatzes}, \& {Lin}}]{bridges1984}
{Bridges}, F.~G., {Hatzes}, A., \& {Lin}, D.~N.~C. 1984, \nat, 309, 333

\bibitem[{{Bromley} \& {Kenyon}(2006)}]{bk2006}
{Bromley}, B.~C., \& {Kenyon}, S.~J. 2006, \aj, 131, 2737

\bibitem[{{Bromley} \& {Kenyon}(2011{\natexlab{a}})}]{bk2011a}
---. 2011{\natexlab{a}}, \apj, 731, 101

\bibitem[{{Bromley} \& {Kenyon}(2011{\natexlab{b}})}]{bk2011b}
---. 2011{\natexlab{b}}, \apj, 735, 29

\bibitem[{{Bromley} \& {Kenyon}(2013)}]{bk2013}
---. 2013, \apj, 764, 192

\bibitem[{{Bromley} \& {Kenyon}(2015)}]{bk2015tatooine}
---. 2015, ArXiv e-prints, arXiv:1503.03876

\bibitem[{{Brozovi{\'c}} {et~al.}(2015){Brozovi{\'c}}, {Showalter}, {Jacobson},
  \& {Buie}}]{brozovic2015}
{Brozovi{\'c}}, M., {Showalter}, M.~R., {Jacobson}, R.~A., \& {Buie}, M.~W.
  2015, \icarus, 246, 317

\bibitem[{{Brucker} {et~al.}(2009){Brucker}, {Grundy}, {Stansberry}, {Spencer},
  {Sheppard}, {Chiang}, \& {Buie}}]{brucker2009}
{Brucker}, M.~J., {Grundy}, W.~M., {Stansberry}, J.~A., {et~al.} 2009, \icarus,
  201, 284

\bibitem[{{Buie} {et~al.}(2013){Buie}, {Grundy}, \& {Tholen}}]{buie2013}
{Buie}, M.~W., {Grundy}, W.~M., \& {Tholen}, D.~J. 2013, \aj, 146, 152

\bibitem[{{Buie} {et~al.}(2006){Buie}, {Grundy}, {Young}, {Young}, \&
  {Stern}}]{buie2006}
{Buie}, M.~W., {Grundy}, W.~M., {Young}, E.~F., {Young}, L.~A., \& {Stern},
  S.~A. 2006, \aj, 132, 290

\bibitem[{{Buie} {et~al.}(2012){Buie}, {Tholen}, \& {Grundy}}]{buie2012}
{Buie}, M.~W., {Tholen}, D.~J., \& {Grundy}, W.~M. 2012, \aj, 144, 15

\bibitem[{{Canup}(2005)}]{canup2005}
{Canup}, R.~M. 2005, Science, 307, 546

\bibitem[{{Canup}(2011)}]{canup2011}
---. 2011, \aj, 141, 35

\bibitem[{{Canup} \& {Esposito}(1995)}]{canup1995}
{Canup}, R.~M., \& {Esposito}, L.~W. 1995, \icarus, 113, 331

\bibitem[{{Canup} \& {Ward}(2002)}]{canup2002}
{Canup}, R.~M., \& {Ward}, W.~R. 2002, \aj, 124, 3404

\bibitem[{{Charnoz} {et~al.}(2009){Charnoz}, {Dones}, {Esposito}, {Estrada}, \&
  {Hedman}}]{charnoz2009}
{Charnoz}, S., {Dones}, L., {Esposito}, L.~W., {Estrada}, P.~R., \& {Hedman},
  M.~M. 2009, in Saturn from Cassini-Huygens, ed. M.~K. {Dougherty}, L.~W.
  {Esposito}, \& S.~M. {Krimigis} (Dordrecht: Springer Science \& Business
  Media), 537

\bibitem[{{Charnoz} {et~al.}(2010){Charnoz}, {Salmon}, \&
  {Crida}}]{charnoz2010}
{Charnoz}, S., {Salmon}, J., \& {Crida}, A. 2010, \nat, 465, 752

\bibitem[{{Charnoz} {et~al.}(2011){Charnoz}, {Crida}, {Castillo-Rogez},
  {Lainey}, {Dones}, {Karatekin}, {Tobie}, {Mathis}, {Le Poncin-Lafitte}, \&
  {Salmon}}]{charnoz2011}
{Charnoz}, S., {Crida}, A., {Castillo-Rogez}, J.~C., {et~al.} 2011, \icarus,
  216, 535

\bibitem[{{Cheng} {et~al.}(2014{\natexlab{a}}){Cheng}, {Lee}, \&
  {Peale}}]{cheng2014a}
{Cheng}, W.~H., {Lee}, M.~H., \& {Peale}, S.~J. 2014{\natexlab{a}}, \icarus,
  233, 242

\bibitem[{{Cheng} {et~al.}(2014{\natexlab{b}}){Cheng}, {Peale}, \&
  {Lee}}]{cheng2014b}
{Cheng}, W.~H., {Peale}, S.~J., \& {Lee}, M.~H. 2014{\natexlab{b}}, \icarus,
  241, 180

\bibitem[{{Chiang} \& {Youdin}(2010)}]{chiang2010}
{Chiang}, E., \& {Youdin}, A.~N. 2010, Annual Review of Earth and Planetary
  Sciences, 38, 493

\bibitem[{{Christy} \& {Harrington}(1978)}]{christy1978}
{Christy}, J.~W., \& {Harrington}, R.~S. 1978, \aj, 83, 1005

\bibitem[{{Colwell} {et~al.}(2009){Colwell}, {Nicholson}, {Tiscareno},
  {Murray}, {French}, \& {Marouf}}]{colwell2009}
{Colwell}, J.~E., {Nicholson}, P.~D., {Tiscareno}, M.~S., {et~al.} 2009, in
  Saturn from Cassini-Huygens, ed. M.~K. {Dougherty}, L.~W. {Esposito}, \&
  S.~M. {Krimigis}, 375

\bibitem[{{Cook} \& {Franklin}(1964)}]{cook1964}
{Cook}, A.~F., \& {Franklin}, F.~A. 1964, \aj, 69, 173

\bibitem[{{Dobrovolskis} {et~al.}(1997){Dobrovolskis}, {Peale}, \&
  {Harris}}]{dobro1997}
{Dobrovolskis}, A.~R., {Peale}, S.~J., \& {Harris}, A.~W. 1997, in Pluto and
  Charon, ed. S.~A. {Stern} \& D.~J. {Tholen}, 159

\bibitem[{{Doolin} \& {Blundell}(2011)}]{doolin2011}
{Doolin}, S., \& {Blundell}, K.~M. 2011, \mnras, 418, 2656

\bibitem[{{Doyle} {et~al.}(2011){Doyle}, {Carter}, {Fabrycky}, {Slawson},
  {Howell}, {Winn}, {Orosz}, {Prsa}, {Welsh}, {Quinn}, {Latham}, {Torres},
  {Buchhave}, {Marcy}, {Fortney}, {Shporer}, {Ford}, {Lissauer}, {Ragozzine},
  {Rucker}, {Batalha}, {Jenkins}, {Borucki}, {Koch}, {Middour}, {Hall},
  {McCauliff}, {Fanelli}, {Quintana}, {Holman}, {Caldwell}, {Still},
  {Stefanik}, {Brown}, {Esquerdo}, {Tang}, {Furesz}, {Geary}, {Berlind},
  {Calkins}, {Short}, {Steffen}, {Sasselov}, {Dunham}, {Cochran}, {Boss},
  {Haas}, {Buzasi}, \& {Fischer}}]{doyle2011}
{Doyle}, L.~R., {Carter}, J.~A., {Fabrycky}, D.~C., {et~al.} 2011, Science,
  333, 1602

\bibitem[{{Farinella} {et~al.}(1979){Farinella}, {Milani}, {Nobili}, \&
  {Valsecchi}}]{farinella1979}
{Farinella}, P., {Milani}, A., {Nobili}, A.~M., \& {Valsecchi}, G.~B. 1979,
  Moon and Planets, 20, 415

\bibitem[{{Georgakarakos} \& {Eggl}(2015)}]{georgakarakos2015}
{Georgakarakos}, N., \& {Eggl}, S. 2015, \apj, 802, 94

\bibitem[{{Gladman}(1993)}]{glad1993}
{Gladman}, B. 1993, \icarus, 106, 247

\bibitem[{{Glaschke} {et~al.}(2014){Glaschke}, {Amaro-Seoane}, \&
  {Spurzem}}]{glaschke2014}
{Glaschke}, P., {Amaro-Seoane}, P., \& {Spurzem}, R. 2014, \mnras, 445, 3620

\bibitem[{{Goldreich} {et~al.}(2004){Goldreich}, {Lithwick}, \&
  {Sari}}]{gold2004}
{Goldreich}, P., {Lithwick}, Y., \& {Sari}, R. 2004, \araa, 42, 549

\bibitem[{{Goldreich} \& {Tremaine}(1980)}]{gold1980}
{Goldreich}, P., \& {Tremaine}, S. 1980, \apj, 241, 425

\bibitem[{{Goldreich} \& {Tremaine}(1982)}]{gold1982}
---. 1982, \araa, 20, 249

\bibitem[{{Goldreich} \& {Tremaine}(1978)}]{gold1978}
{Goldreich}, P., \& {Tremaine}, S.~D. 1978, Icarus, 34, 227

\bibitem[{{Greenzweig} \& {Lissauer}(1990)}]{green1990}
{Greenzweig}, Y., \& {Lissauer}, J.~J. 1990, Icarus, 87, 40

\bibitem[{{Heppenheimer}(1978)}]{hepp1978}
{Heppenheimer}, T.~A. 1978, \aap, 65, 421

\bibitem[{{Holman} \& {Wiegert}(1999)}]{holman1999}
{Holman}, M.~J., \& {Wiegert}, P.~A. 1999, \aj, 117, 621

\bibitem[{{Holsapple} \& {Michel}(2006)}]{holsapple2006}
{Holsapple}, K.~A., \& {Michel}, P. 2006, \icarus, 183, 331

\bibitem[{{Holsapple} \& {Michel}(2008)}]{holsapple2008}
---. 2008, \icarus, 193, 283

\bibitem[{{Hornung} {et~al.}(1985){Hornung}, {Pellat}, \& {Barge}}]{horn1985}
{Hornung}, P., {Pellat}, R., \& {Barge}, P. 1985, Icarus, 64, 295

\bibitem[{{Ida} {et~al.}(2000){Ida}, {Bryden}, {Lin}, \& {Tanaka}}]{ida2000b}
{Ida}, S., {Bryden}, G., {Lin}, D.~N.~C., \& {Tanaka}, H. 2000, \apj, 534, 428

\bibitem[{{Ida} \& {Makino}(1993)}]{ida1993}
{Ida}, S., \& {Makino}, J. 1993, \icarus, 106, 210

\bibitem[{{Jeffreys}(1947)}]{jeffreys1947}
{Jeffreys}, H. 1947, \mnras, 107, 263

\bibitem[{{Kenyon} \& {Bromley}(2001)}]{kb2001}
{Kenyon}, S.~J., \& {Bromley}, B.~C. 2001, \aj, 121, 538

\bibitem[{{Kenyon} \& {Bromley}(2004)}]{kb2004a}
---. 2004, \aj, 127, 513

\bibitem[{{Kenyon} \& {Bromley}(2014)}]{kb2014}
---. 2014, \aj, 147, 8

\bibitem[{{Kenyon} {et~al.}(2014){Kenyon}, {Currie}, \& {Bromley}}]{kcb2014}
{Kenyon}, S.~J., {Currie}, T., \& {Bromley}, B.~C. 2014, \apj, 786, 70

\bibitem[{{Kenyon} \& {Luu}(1998)}]{kl1998}
{Kenyon}, S.~J., \& {Luu}, J.~X. 1998, \aj, 115, 2136

\bibitem[{{Lecar} {et~al.}(2001){Lecar}, {Franklin}, {Holman}, \&
  {Murray}}]{lecar2001}
{Lecar}, M., {Franklin}, F.~A., {Holman}, M.~J., \& {Murray}, N.~J. 2001,
  \araa, 39, 581

\bibitem[{{Lee} \& {Peale}(2006)}]{lee2006}
{Lee}, M.~H., \& {Peale}, S.~J. 2006, \icarus, 184, 573

\bibitem[{{Leinhardt} \& {Stewart}(2012)}]{lein2012}
{Leinhardt}, Z.~M., \& {Stewart}, S.~T. 2012, \apj, 745, 79

\bibitem[{{Leung} \& {Lee}(2013)}]{leung2013}
{Leung}, G.~C.~K., \& {Lee}, M.~H. 2013, \apj, 763, 107

\bibitem[{{Lin} \& {Papaloizou}(1979{\natexlab{a}})}]{lin1979b}
{Lin}, D.~N.~C., \& {Papaloizou}, J. 1979{\natexlab{a}}, \mnras, 188, 191

\bibitem[{{Lin} \& {Papaloizou}(1979{\natexlab{b}})}]{lin1979a}
---. 1979{\natexlab{b}}, \mnras, 186, 799

\bibitem[{{Lines} {et~al.}(2014){Lines}, {Leinhardt}, {Paardekooper},
  {Baruteau}, \& {Thebault}}]{lines2014}
{Lines}, S., {Leinhardt}, Z.~M., {Paardekooper}, S., {Baruteau}, C., \&
  {Thebault}, P. 2014, \apjl, 782, L11

\bibitem[{{Lissauer}(1987)}]{liss1987}
{Lissauer}, J.~J. 1987, Icarus, 69, 249

\bibitem[{{Lithwick} \& {Wu}(2008)}]{lith2008b}
{Lithwick}, Y., \& {Wu}, Y. 2008, ArXiv e-prints, arXiv:0802.2951

\bibitem[{{Lunine} \& {Stevenson}(1982)}]{lun1982}
{Lunine}, J.~I., \& {Stevenson}, D.~J. 1982, \icarus, 52, 14

\bibitem[{{Lynden-Bell} \& {Pringle}(1974)}]{lbp1974}
{Lynden-Bell}, D., \& {Pringle}, J.~E. 1974, \mnras, 168, 603

\bibitem[{{McKinnon}(1989)}]{mckinnon1989}
{McKinnon}, W.~B. 1989, \apjl, 344, L41

\bibitem[{{Meyer-Vernet} \& {Sicardy}(1987)}]{meyer1987}
{Meyer-Vernet}, N., \& {Sicardy}, B. 1987, \icarus, 69, 157

\bibitem[{{Moriwaki} \& {Nakagawa}(2004)}]{mori2004}
{Moriwaki}, K., \& {Nakagawa}, Y. 2004, \apj, 609, 1065

\bibitem[{{Mosqueira} {et~al.}(2010){Mosqueira}, {Estrada}, \&
  {Turrini}}]{mosque2010}
{Mosqueira}, I., {Estrada}, P., \& {Turrini}, D. 2010, \ssr, 153, 431

\bibitem[{{Mosqueira} \& {Estrada}(2003{\natexlab{a}})}]{mosque2003a}
{Mosqueira}, I., \& {Estrada}, P.~R. 2003{\natexlab{a}}, \icarus, 163, 198

\bibitem[{{Mosqueira} \& {Estrada}(2003{\natexlab{b}})}]{mosque2003b}
---. 2003{\natexlab{b}}, \icarus, 163, 232

\bibitem[{{Murray} \& {Dermott}(1999)}]{murray1999}
{Murray}, C.~D., \& {Dermott}, S.~F. 1999, {Solar system dynamics} (Princeton:
  Princeton University Press)

\bibitem[{{Musielak} {et~al.}(2005){Musielak}, {Cuntz}, {Marshall}, \&
  {Stuit}}]{musielak2005}
{Musielak}, Z.~E., {Cuntz}, M., {Marshall}, E.~A., \& {Stuit}, T.~D. 2005,
  \aap, 434, 355

\bibitem[{{Ogihara} \& {Ida}(2012)}]{ogi2012}
{Ogihara}, M., \& {Ida}, S. 2012, \apj, 753, 60

\bibitem[{{Ohtsuki}(1992)}]{oht1992}
{Ohtsuki}, K. 1992, Icarus, 98, 20

\bibitem[{{Ohtsuki}(1999)}]{oht1999}
---. 1999, \icarus, 137, 152

\bibitem[{{Ohtsuki} {et~al.}(2002){Ohtsuki}, {Stewart}, \& {Ida}}]{oht2002}
{Ohtsuki}, K., {Stewart}, G.~R., \& {Ida}, S. 2002, Icarus, 155, 436

\bibitem[{{Ohtsuki} \& {Tanaka}(2003)}]{oht2003}
{Ohtsuki}, K., \& {Tanaka}, H. 2003, \icarus, 162, 47

\bibitem[{{Ormel} {et~al.}(2012){Ormel}, {Ida}, \& {Tanaka}}]{ormel2012}
{Ormel}, C.~W., {Ida}, S., \& {Tanaka}, H. 2012, \apj, 758, 80

\bibitem[{{Orosz} {et~al.}(2012{\natexlab{a}}){Orosz}, {Welsh}, {Carter},
  {Fabrycky}, {Cochran}, {Endl}, {Ford}, {Haghighipour}, {MacQueen}, {Mazeh},
  {Sanchis-Ojeda}, {Short}, {Torres}, {Agol}, {Buchhave}, {Doyle}, {Isaacson},
  {Lissauer}, {Marcy}, {Shporer}, {Windmiller}, {Barclay}, {Boss}, {Clarke},
  {Fortney}, {Geary}, {Holman}, {Huber}, {Jenkins}, {Kinemuchi}, {Kruse},
  {Ragozzine}, {Sasselov}, {Still}, {Tenenbaum}, {Uddin}, {Winn}, {Koch}, \&
  {Borucki}}]{orosz2012a}
{Orosz}, J.~A., {Welsh}, W.~F., {Carter}, J.~A., {et~al.} 2012{\natexlab{a}},
  Science, 337, 1511

\bibitem[{{Orosz} {et~al.}(2012{\natexlab{b}}){Orosz}, {Welsh}, {Carter},
  {Brugamyer}, {Buchhave}, {Cochran}, {Endl}, {Ford}, {MacQueen}, {Short},
  {Torres}, {Windmiller}, {Agol}, {Barclay}, {Caldwell}, {Clarke}, {Doyle},
  {Fabrycky}, {Geary}, {Haghighipour}, {Holman}, {Ibrahim}, {Jenkins},
  {Kinemuchi}, {Li}, {Lissauer}, {Pr{\v s}a}, {Ragozzine}, {Shporer}, {Still},
  \& {Wade}}]{orosz2012b}
---. 2012{\natexlab{b}}, \apj, 758, 87

\bibitem[{{Paardekooper} {et~al.}(2012){Paardekooper}, {Leinhardt},
  {Th{\'e}bault}, \& {Baruteau}}]{paarde2012}
{Paardekooper}, S.-J., {Leinhardt}, Z.~M., {Th{\'e}bault}, P., \& {Baruteau},
  C. 2012, \apjl, 754, L16

\bibitem[{{Peale}(1999)}]{peale1999}
{Peale}, S.~J. 1999, \araa, 37, 533

\bibitem[{{Peale} {et~al.}(2011){Peale}, {Cheng}, \& {Lee}}]{peale2011}
{Peale}, S.~J., {Cheng}, W.~H., \& {Lee}, M.~H. 2011, in EPSC-DPS Joint Meeting
  2011, 665

\bibitem[{{Person} {et~al.}(2006){Person}, {Elliot}, {Gulbis}, {Pasachoff},
  {Babcock}, \& {Souza}}]{person2006}
{Person}, M.~J., {Elliot}, J.~L., {Gulbis}, A.~A.~S., {et~al.} 2006, \aj, 132,
  1575

\bibitem[{{Pichardo} {et~al.}(2005){Pichardo}, {Sparke}, \&
  {Aguilar}}]{pichardo2005}
{Pichardo}, B., {Sparke}, L.~S., \& {Aguilar}, L.~A. 2005, \mnras, 359, 521

\bibitem[{{Pichardo} {et~al.}(2008){Pichardo}, {Sparke}, \&
  {Aguilar}}]{pichardo2008}
---. 2008, \mnras, 391, 815

\bibitem[{{Pierens} \& {Nelson}(2007)}]{pierens2007}
{Pierens}, A., \& {Nelson}, R.~P. 2007, \aap, 472, 993

\bibitem[{{Popova} \& {Shevchenko}(2013)}]{popova2013}
{Popova}, E.~A., \& {Shevchenko}, I.~I. 2013, \apj, 769, 152

\bibitem[{{Porco} {et~al.}(2007){Porco}, {Thomas}, {Weiss}, \&
  {Richardson}}]{porco2007}
{Porco}, C.~C., {Thomas}, P.~C., {Weiss}, J.~W., \& {Richardson}, D.~C. 2007,
  Science, 318, 1602

\bibitem[{{Porco} {et~al.}(2008){Porco}, {Weiss}, {Richardson}, {Dones},
  {Quinn}, \& {Throop}}]{porco2008}
{Porco}, C.~C., {Weiss}, J.~W., {Richardson}, D.~C., {et~al.} 2008, \aj, 136,
  2172

\bibitem[{{Pringle}(1981)}]{pri1981}
{Pringle}, J.~E. 1981, \araa, 19, 137

\bibitem[{{Pringle}(1991)}]{pri1991}
---. 1991, \mnras, 248, 754

\bibitem[{{Quintana} \& {Lissauer}(2006)}]{quintana2006}
{Quintana}, E.~V., \& {Lissauer}, J.~J. 2006, \icarus, 185, 1

\bibitem[{{Rafikov}(2013)}]{raf2013}
{Rafikov}, R.~R. 2013, \apjl, 764, L16

\bibitem[{{Rosenblatt} \& {Charnoz}(2012)}]{rosen2012}
{Rosenblatt}, P., \& {Charnoz}, S. 2012, \icarus, 221, 806

\bibitem[{{Safronov}(1969)}]{saf1969}
{Safronov}, V.~S. 1969, {Evoliutsiia doplanetnogo oblaka. (Evolution of the
  Protoplanetary Cloud and Formation of the Earth and Planets, Nauka, Moscow
  [Translation 1972, NASA TT F-677]} (1969.)

\bibitem[{{Salmon} {et~al.}(2010){Salmon}, {Charnoz}, {Crida}, \&
  {Brahic}}]{salmon2010}
{Salmon}, J., {Charnoz}, S., {Crida}, A., \& {Brahic}, A. 2010, \icarus, 209,
  771

\bibitem[{{Sasaki} {et~al.}(2010){Sasaki}, {Stewart}, \& {Ida}}]{sas2010}
{Sasaki}, T., {Stewart}, G.~R., \& {Ida}, S. 2010, \apj, 714, 1052

\bibitem[{{Showalter} {et~al.}(2013){Showalter}, {Weaver}, {Buie}, {Merline},
  {Mutchler}, {Soummer}, {Steffl}, {Stern}, {Throop}, \&
  {Young}}]{showalter2013}
{Showalter}, M., {Weaver}, H., {Buie}, M., {et~al.} 2013, in EGU General
  Assembly Conference Abstracts, Vol.~15, EGU General Assembly Conference
  Abstracts, 13786

\bibitem[{{Showalter} \& {Hamilton}(2015)}]{showalter2015}
{Showalter}, M.~R., \& {Hamilton}, D.~P. 2015, \nat, 522, 45

\bibitem[{{Showalter} {et~al.}(2011){Showalter}, {Hamilton}, {Stern}, {Weaver},
  {Steffl}, \& {Young}}]{showalter2011}
{Showalter}, M.~R., {Hamilton}, D.~P., {Stern}, S.~A., {et~al.} 2011, \iaucirc,
  9221, 1

\bibitem[{{Showalter} {et~al.}(2012){Showalter}, {Weaver}, {Stern}, {Steffl},
  {Buie}, {Merline}, {Mutchler}, {Soummer}, \& {Throop}}]{showalter2012}
{Showalter}, M.~R., {Weaver}, H.~A., {Stern}, S.~A., {et~al.} 2012, \iaucirc,
  9253, 1

\bibitem[{{Shu} \& {Stewart}(1985)}]{shu1985}
{Shu}, F.~H., \& {Stewart}, G.~R. 1985, \icarus, 62, 360

\bibitem[{{Spaute} {et~al.}(1991){Spaute}, {Weidenschilling}, {Davis}, \&
  {Marzari}}]{spaute1991}
{Spaute}, D., {Weidenschilling}, S.~J., {Davis}, D.~R., \& {Marzari}, F. 1991,
  Icarus, 92, 147

\bibitem[{{Stern}(1992)}]{stern1992}
{Stern}, S.~A. 1992, \araa, 30, 185

\bibitem[{{Stern}(2008)}]{stern2008}
---. 2008, \ssr, 140, 3

\bibitem[{{Stern} {et~al.}(2006){Stern}, {Weaver}, {Steffl}, {Mutchler},
  {Merline}, {Buie}, {Young}, {Young}, \& {Spencer}}]{stern2006}
{Stern}, S.~A., {Weaver}, H.~A., {Steffl}, A.~J., {et~al.} 2006, \nat, 439, 946

\bibitem[{{Stewart} \& {Ida}(2000)}]{stewart:2000}
{Stewart}, G.~R., \& {Ida}, S. 2000, Icarus, 143, 28

\bibitem[{{S{\"u}li} \& {Zsigmond}(2009)}]{suli2009}
{S{\"u}li}, {\'A}., \& {Zsigmond}, Z. 2009, \mnras, 398, 2199

\bibitem[{{Supulver} {et~al.}(1995){Supulver}, {Bridges}, \& {Lin}}]{sup1995}
{Supulver}, K.~D., {Bridges}, F.~G., \& {Lin}, D.~N.~C. 1995, \icarus, 113, 188

\bibitem[{{Tholen} {et~al.}(2012){Tholen}, {Buie}, \& {Grundy}}]{tholen2012}
{Tholen}, D.~J., {Buie}, M.~W., \& {Grundy}, W.~M. 2012, LPI Contributions,
  1667, 6327

\bibitem[{{Tholen} {et~al.}(2008){Tholen}, {Buie}, {Grundy}, \&
  {Elliott}}]{tholen2008}
{Tholen}, D.~J., {Buie}, M.~W., {Grundy}, W.~M., \& {Elliott}, G.~T. 2008, \aj,
  135, 777

\bibitem[{{Toomre}(1964)}]{toom64}
{Toomre}, A. 1964, \apj, 139, 1217

\bibitem[{{Walsh} \& {Levison}(2015)}]{walsh2015}
{Walsh}, K.~J., \& {Levison}, H.~F. 2015, ArXiv e-prints, arXiv:1505.01208

\bibitem[{{Ward}(1997)}]{ward1997}
{Ward}, W.~R. 1997, \icarus, 126, 261

\bibitem[{{Ward} \& {Canup}(2006)}]{ward2006}
{Ward}, W.~R., \& {Canup}, R.~M. 2006, Science, 313, 1107

\bibitem[{{Ward} \& {Canup}(2010)}]{ward2010}
---. 2010, \aj, 140, 1168

\bibitem[{{Weaver} {et~al.}(2006){Weaver}, {Stern}, {Mutchler}, {Steffl},
  {Buie}, {Merline}, {Spencer}, {Young}, \& {Young}}]{weaver2006}
{Weaver}, H.~A., {Stern}, S.~A., {Mutchler}, M.~J., {et~al.} 2006, \nat, 439,
  943

\bibitem[{{Weidenschilling}(1989)}]{weiden1989}
{Weidenschilling}, S.~J. 1989, Icarus, 80, 179

\bibitem[{{Weidenschilling} {et~al.}(1997){Weidenschilling}, {Spaute}, {Davis},
  {Marzari}, \& {Ohtsuki}}]{weiden1997b}
{Weidenschilling}, S.~J., {Spaute}, D., {Davis}, D.~R., {Marzari}, F., \&
  {Ohtsuki}, K. 1997, Icarus, 128, 429

\bibitem[{{Welsh} {et~al.}(2012){Welsh}, {Orosz}, {Carter}, {Fabrycky}, {Ford},
  {Lissauer}, {Pr{\v s}a}, {Quinn}, {Ragozzine}, {Short}, {Torres}, {Winn},
  {Doyle}, {Barclay}, {Batalha}, {Bloemen}, {Brugamyer}, {Buchhave},
  {Caldwell}, {Caldwell}, {Christiansen}, {Ciardi}, {Cochran}, {Endl},
  {Fortney}, {Gautier}, {Gilliland}, {Haas}, {Hall}, {Holman}, {Howard},
  {Howell}, {Isaacson}, {Jenkins}, {Klaus}, {Latham}, {Li}, {Marcy}, {Mazeh},
  {Quintana}, {Robertson}, {Shporer}, {Steffen}, {Windmiller}, {Koch}, \&
  {Borucki}}]{welsh2012}
{Welsh}, W.~F., {Orosz}, J.~A., {Carter}, J.~A., {et~al.} 2012, \nat, 481, 475

\bibitem[{{Wetherill} \& {Cox}(1985)}]{weth1985a}
{Wetherill}, G.~W., \& {Cox}, L.~P. 1985, Icarus, 63, 290

\bibitem[{{Wetherill} \& {Stewart}(1989)}]{weth1989}
{Wetherill}, G.~W., \& {Stewart}, G.~R. 1989, Icarus, 77, 330

\bibitem[{{Wetherill} \& {Stewart}(1993)}]{weth1993}
---. 1993, Icarus, 106, 190

\bibitem[{{Winter} {et~al.}(2010){Winter}, {Winter}, {Guimar{\~a}es}, \&
  {Silva}}]{winter2010}
{Winter}, S.~M.~G., {Winter}, O.~C., {Guimar{\~a}es}, A.~H.~F., \& {Silva},
  M.~R. 2010, \mnras, 404, 442

\bibitem[{{Wisdom}(1980)}]{wisdom1980}
{Wisdom}, J. 1980, \aj, 85, 1122

\bibitem[{{Wyatt} {et~al.}(1999){Wyatt}, {Dermott}, {Telesco}, {Fisher},
  {Grogan}, {Holmes}, \& {Pi{\~n}a}}]{wyatt1999}
{Wyatt}, M.~C., {Dermott}, S.~F., {Telesco}, C.~M., {et~al.} 1999, \apj, 527,
  918

\bibitem[{{Youdin} \& {Kenyon}(2013)}]{youdin2013}
{Youdin}, A.~N., \& {Kenyon}, S.~J. 2013, in Planets, Stars and Stellar
  Systems.~Volume 3: Solar and Stellar Planetary Systems, ed. T.~D. {Oswalt},
  L.~M. {French}, \& P.~{Kalas} (Dordrecht: Springer Science \& Business
  Media), 1

\bibitem[{{Youdin} {et~al.}(2012){Youdin}, {Kratter}, \& {Kenyon}}]{youdin2012}
{Youdin}, A.~N., {Kratter}, K.~M., \& {Kenyon}, S.~J. 2012, \apj, 755, 17

\bibitem[{{Young} \& {Binzel}(1994)}]{young1994}
{Young}, E.~F., \& {Binzel}, R.~P. 1994, \icarus, 108, 219

\bibitem[{{Young} {et~al.}(2007){Young}, {Young}, \& {Buie}}]{young2007}
{Young}, E.~F., {Young}, L.~A., \& {Buie}, M. 2007, in Bulletin of the American
  Astronomical Society, Vol.~39, AAS/Division for Planetary Sciences Meeting
  Abstracts \#39, 541

\end{thebibliography}
\bibliographystyle{apj}


\end{document}